\documentclass[prb,twocolumn,floatfix,showpacs]{revtex4-1}
\usepackage{epsfig,subfigure,amssymb,amscd,amsmath,graphicx,amsfonts,mathtools,mathcomp,pbox}
\usepackage{times,longtable,setspace,fancyhdr,flushend,tabularx}

\newcommand{\non}{\nonumber}

\newcommand{\bea}{\begin{eqnarray}}
\newcommand{\eea}{\end{eqnarray}}
\newcommand{\be}{\begin{equation}}
\newcommand{\ee}{\end{equation}}
\newcommand{\ba}{\begin{align}}
\newcommand{\ea}{\end{align}}
\newcommand{\braket}[2]{\langle #1|#2\rangle}
\newcommand{\ket}[1]{     |    \,    #1    \rangle}
\newcommand{\bra}[1]{  \langle #1  \,  |}

\newcommand{\phd}{ {\phantom\dagger} }

\usepackage[usenames]{color}

\DeclareMathOperator{\Tr}{Tr}
\makeatletter
\newcommand*\dashline{\rotatebox[origin=c]{90}{$\dabar@\dabar@\dabar@$}}
\makeatother

\newcommand{\Ket}[1]{     \dashline    \,    #1    \big)}
\newcommand{\BraKet}[2]{\big( #1 \dashline #2\big)}
\newcommand{\Bra}[1]{  \big( #1  \,  \dashline}

\begin{document}

\title{Multi-particle content of Majorana zero-modes in the interacting p-wave wire.}

\author{G. Kells }

\affiliation{ Dublin Institute for Advanced  Studies, School of Theoretical Physics, 10 Burlington Rd, Dublin 4, Ireland.}

\begin{abstract}
In the topological phase of p-wave superconductors, zero-energy Majorana quasi-particle excitations can be well-defined in the presence of local density-density interactions.  Here we examine this phenomenon from the perspective of matrix representations of the commutator $\mathcal{H} =[H,\bullet]$ ,with the aim of characterising the multi-particle content of the many-body Majorana mode.   To do this we  show that, for quadratic fermionic systems, $\mathcal{H}$  can always be decomposed into sub-blocks that act as multi-particle generalisations of the BdG/Majorana forms that encode single-particle excitations. In this picture, density-density like interactions will break this exact excitation-number symmetry, coupling different sub-blocks and lifting degeneracies so that the eigen-operators of the commutator $\mathcal{H}$ take the form of individual eigenstate transitions $\ket{n}\bra{m}$. However, the Majorana mode is special in that zero-energy transitions are not destroyed by local interactions and it becomes possible to define many-body Majoranas as the odd-parity zero-energy solutions of $\mathcal{H}$ that minimise their excitation number.  This idea forms the basis for an algorithm which is used to characterise the multi-particle excitation content of the Majorana zero modes of the one-dimensional p-wave lattice model. We find that the multi-particle content of the Majorana zero-mode operators is significant even at modest interaction strengths. This has important consequences for the stability of Majorana based qubits when they are coupled to a heat bath. We will also discuss how these findings differ from previous work regarding the structure of the many-body-Majorana operators and point out that this should affect how certain experimental features are interpreted. 
\end{abstract}

\pacs{74.78.Na  74.20.Rp  03.67.Lx  73.63.Nm}

\date{\today} \maketitle
\thispagestyle{fancy}

\section{Introduction}
Zero-energy Majorana quasi-particles are expected to be pinned to defects and/or domain walls in topological superconductors. \cite{Read2000,Kitaev2001}  These particles are predicted to display fractionalised non-abelian statistics, which may allow for the manipulation of quantum information in a robust manner using non-local braiding operations. \cite{Ivanov2001,Nayak2008, Alicea2011}  There are now a number of potential systems in which these Majorana modes could  potentially be observed, \cite{Alicea2012,Beenakker2013,Stanescu2013} the most well-known being those based on proximity-coupled semiconductor nano-wires. \cite{Oreg2010,Lutchyn2010} In these nano-wire systems, observations of  anomalous zero-bias conductances are a strong experimental indication of the Majorana modes.\cite{Mourik2012,Das2012,Churchill2013}  More  recently, alternative approaches using magnetic molecules, whose bound states can be resolved energetically and spatially, have also attracted interest.\cite{Nadj-Perge2014}

On a mean-field level, the notion of Majorana quasi-particle has proved enormously useful as both a conceptual and calculational tool. There is therefore  ample reason to explore how much of this quasi-particle picture remains valid beyond the confines of  mean-field superconductivity.  For example,  considerable progress has been made towards developing number preserving theories of the Majorana modes,\cite{Fidkowski2011,Sau2011,Cheng2011,Kraus2013,Ortiz2014,Iemini2015}  as well as a growing body of work which examines how  free-topological superconducting phases are affected by the addition of interacting electron-electron terms. \cite{Lutchyn2011,Sela2011,Stoudenmire2011,Gangadharaiah2011,Lobos2012,Hassler2012,Goldstein2012,Yang2014,Thomale2013,Sticlet2014,Kells2014,Katsura2015,Chan2015}  One aspect of this latter story is concerned with the stability and structure of the Majorana zero-modes themselves and how they are affected by the presence of density-density interaction terms that break the exactly solvable nature of the underlying model.  The issue of stable zero-modes has also been addressed in the related context of 1-d parafermionic chains.  \cite{Fendley2012, Fendley2013, Jermyn2014}  

To make the following discussion precise,  note that in the topological phase of the 1-dimensional p-wave superconducting wires the Majorana modes/operators are exponentially localised at each end of the wire.\cite{Kitaev2001} In the long-wire limit ($N\rightarrow \infty$) the (L)eft and (R)ight Majorana modes have precisely the energy $E=0$ and the corresponding operators can have the form
\bea \label{eq:Majdef}
\gamma_L  &=&  \sum_i^{N}  i (c^\dagger_i-c_i) u_L(i) \\ \non
\gamma_R  &=&  \phantom{i} \sum_i^{N}  (c^\dagger_i+c_i) u_R(i) 
\eea
where  $u_L(i)$ and $u_R(i)$ are the single particle wave-functions localised to the left and right of the wire, and the $c^\dagger$'s and $c$'s are  the Dirac fermion creation and  annihilation operators respectively. For free-fermionic systems the existence of these zero-energy solutions can be easily established throughout the topological region.\cite{Kitaev2001}  It is important to note that the mode stability has nothing to do with any rigidity in the form of the functions $u_L$ and $u_R$.  Indeed, the functions themselves are actually very susceptible to variations in the underlying system and it is this fluidity that allows the zero-energy Majorana modes to exist even in highly disordered regions of the topological phase, see for example Ref. \onlinecite{Rieder2013}.

The zero-mode operators $\gamma_L$ and $\gamma_R$  commute with the Hamiltonian $H$, are self adjoint, and anti-commute with both parity $P$ and each other:
\bea 
[H,\gamma_{L/R}]&=&0, \quad \gamma_{L/R}^2=I, \\ \non
\{ P, \gamma_{L/R} \} &=&0, \quad  \{ \gamma_L, \gamma_R \}= 0 .
\eea
 When interactions are included, and if they exist,  the zero-energy Majorana modes should obey the same criteria but should now appear in the form of  multi-nomial sums:\cite{Goldstein2012}
\bea
\label{eq:Majdef2}
\gamma_L = \sum_i^{2N} u^{(1)}_L(i) \gamma_i + \sum_{ijk}^{2N_x} u^{(3)}_L(i,j,k) \gamma_{i} \gamma_{k} \gamma_{k} + ... \\
\gamma_R = \sum_i^{2N} u^{(1)}_R(i) \gamma_i + \sum_{ijk}^{2N_x} u^{(3)}_R(i,j,k) \gamma_{i}  \gamma_{j} \gamma_{k} + ... \non
\eea
where $\gamma_{2i-1} = c^\dagger_i+c_i$ and $\gamma_{2i} = i(c_i^\dagger -c_i)$.   However, establishing the existence of Many-Body-Majorana (MBM)-modes outside of the context of mean-field superconductivity has not been straightforward.  Refs. \onlinecite{Goldstein2012} and \onlinecite{Yang2014} addressed this issue on the level of general Hamiltonians and showed that interacting zero-modes of the type above can always be well defined in the presence of local parity-preserving interacting terms, provided  there are only an odd number of  participating Majorana modes. These papers  also show that the notion of a zero-mode can survive when the system is coupled to additional bosonic degrees of freedom.  This in turn can be used to describe in what way the generalised parity based qubits  (see e.g. Ref. \onlinecite{Akhmerov2010}) are susceptible to thermal noise.  

The question of well-defined MBM-modes was also addressed in the specific context of one-dimensional  wires. In  Ref. \onlinecite{Gangadharaiah2011}  it was shown that, when the p-wave system can be bosonized,  a refermionization argument indicates the continued stability of the modes in interacting regions of the topological phase.  Importantly, this argument does not require the restriction to an odd number of Majorana modes.   The re-fermionization procedure casts the many-body Majorana operator in a form that resembles a renormalised single-particle wavefunction of the form Eq. \eqref{eq:Majdef}. Although this allows us to examine general features that the operator in a single-particle picture, it does not imply that the many-body contributions to the operator ($u^{(3)}$, $u^{(5)}$ etc.)  are suppressed.  This is an important point because in related work on the proximity coupled semi-conductor model,\cite{Stoudenmire2011} calculations of the weights of the linear ground-state cross-correlators, obtained using DMRG/MPS techniques, do actually indicate that the many-body Majorana operators resemble renormalised non-interacting modes, even in the presence of strong interactions.  

The existence of MBM's was established more generally in Ref. \onlinecite{Kells2014} where it was shown that in the long-wire limit of the Kitaev chain model, when in the topological phase, all eigenstates come in degenerate pairs even in the presence of local interactions.  A general definition of  the many-body Majorana operators follows:
\bea
\label{eq:MBM}
\gamma_R = \phantom{i} \sum_n \ket{n_e} \bra{n_o}+\ket{n_o} \bra{n_e} \\
\gamma_L = i \sum_n \ket{n_e} \bra{n_o}-\ket{n_o} \bra{n_e} \non
\eea
where the states $\ket{n_o}$ and $\ket{n_o}$ are the odd and even eigenstates of the Hamiltonian.  This result implies that the many-body-Majorana modes can be well-defined in the topological region without the {\em a priori} restriction on the numbers of the participating Majorana modes in the defining Hamiltonian, or the requirement that the chemical potential be far from the bottom of the band.  

In this current paper we set out to characterise the many-body zero modes of the generic 1-d p-wave interacting model by numerically calculating the weights 
\be
|N^\Gamma_n|^2  = \int  |u_{L/R}^{(n)} (\vec{x})|^2 d\vec{x}
\ee
where the weights have the property that $\sum |N^\Gamma_n|^2 =1$.   Instead of essentially single particle structure found in Ref.  \onlinecite{Stoudenmire2011}, we find that higher N-particle terms  grow quickly as one increases the interaction strength $U$.  More specifically we find that
\bea
\label{eq:rates}
|N^\Gamma_3( U)| &\propto&  \alpha_3 U   \\
|N^\Gamma_5(U)| &\propto &  \alpha_5 U^{2} \non \\
|N^\Gamma_7(U)| &\propto &  \alpha_7 U^{3} \non \\
\vdots \non
\eea

These scaling rates have direct consequences for topological-quantum-memories. This is because there is a clear link between the multi-particle content of the MBM zero-modes and the rate of decoherence of Majorana-based qubits when they are connected to a heat bath. \cite{Goldstein2012,Yang2014}   The longest decoherence times occur for those modes that minimise their multi-particle content and we shall see that these optimal cases can also be identified with Eq. \eqref{eq:MBM} . This implies that the scaling rates above represent a best-case-scenario, and suggests that Majorana-based qubits in interacting systems are never fully immune to this type of decoherence.
 
We will discuss in detail later why the general scaling results above appear to disagree with Ref. \onlinecite{Stoudenmire2011}. However we will see that it largely follows from different operational definitions of the Majorana quasi-particles in the interacting regime.  While we are interested in computing the unique operators $\gamma_{L/R}$ which take every eigenstate to its degenerate parity-swapped counterpart, the approach of Ref. \onlinecite{Stoudenmire2011} uses a less restrictive notion, where one seeks to quantify how well the two degenerate ground-states may be mapped into each other using the set of single particle operators. More specifically  Ref. \onlinecite{Stoudenmire2011} calculates 
\be
|N^{\text{gs}}_1|^2  = \int  |O_x|^2 dx,
\ee
where $O(x)= \phantom|_e\bra{0} c^\dagger_x \pm c_x \ket{0}_o$. This measure tends to stay very close to unity even at extremely high interaction strengths but  we shall see that, aside from the non-interacting regime, this approach does not allow one to uniquely determine the single particle content of the Majorana mode.  Indeed, the measure $O_x$ generically includes contributions from the higher N-particle parts of the Majorana operator, and hence it is not a reliable measure to use in this specific instance.  

This aspect of the story has relevance to the ongoing discussion on the relative merits of what are called weak and strong zero-modes.\cite{Alicea2015,Alexandradinata2015,OBrien2015}  The results presented here show that when interactions are present, it is typically impossible to infer the structure of the unique strong zero-modes from the properties of the ground-state manifold.  We shall discuss in our concluding remarks why this has implications for how we should interpret experimental measurements, which typically only examine these low-energy states.  
\vspace{3mm}

The paper is structured as follows.  In section \ref{sect:Opreps}  we show how to arrive at operator representations of the commutator $[H,\bullet]$. This discussion clarifies the relationship between commutator approach of Refs. \onlinecite{Goldstein2012,Yang2014} and the degeneracy methodology of Ref. \onlinecite{Kells2014}. In \ref{sect:Majorana_basis} we demonstrate that in the representation generated by the position space Majorana operators,  the non-interacting model commutator $[H,\bullet]$ naturally decomposes into blocks spanned by fermionic transitions of the same number.   In section \ref{sect:Quadratic} we outline how one can understand the relationship between single-particle excitations and the individual energy transitions at the commutator level.  In section \ref{sect:Symmetries} we build on this idea and show how to relate the notion of integrability in quadratic systems and hopping and pairing symmetries in a fermionic model made from two copies of the original. There are similarities between this approach and the iterative methods to find parfermionic zero-modes (see for example  Ref. \onlinecite{Fendley2012}).    In section \ref{sect:SpinKitaev} we give a spin-representation for the Kitaev chain commutator, in which the basis states are transparently related to the differences in fermionic occupation. 

The results of section \ref{sect:Opreps_all} are also used to show that the MBM quasi-particles  (\ref{eq:MBM})  are the odd parity zero-modes that minimise their total excitation number.  This latter idea forms the basis for an algorithm outlined in section \ref{sect:Algorithms} to calculate the MBM solutions for larger system sizes than are possible with the exact diagonalization method of Ref. \onlinecite{Kells2014}. In section \ref{sect:Numbers} we discuss the resulting numerical data, focusing in particular on the N-particle participation rates mentioned in Eq. \eqref{eq:rates}. We also present additional analysis of the how the multi-particle content grow with respect to the other parameters of the model noting in particular  the clear parabolic dependence of $\alpha_n$ around $\mu=2t$, where were have linear dispersion. In section \ref{sect:correlators} we outline why ground state correlators will necessarily underestimate the many-particle contributions to the Majorana zero-mode in the presence of interactions.

We also include a number of appendices to help make this paper more self-contained. In Appendix  \ref{sect:Kitaevchain} we briefly review the p-wave Hamiltonian. In Appendix   \ref{sect:spinGamma} we show how to derive a spin-representation using the algebra of position space Majorana  operators.  This basis is naturally block diagonal for quadratic Hamiltonians.  In Appendix   \ref{sect:spinSigma}  we show how to derive a spin representation using the algebra of position space fermioninc creation and annihilation operators. This basis shows us how the commutator $[H,\bullet]$ can be thought of as a doubled fermionic system and how the symmetry responsible for the aforementioned block decomposition can be understood as fermionic hoppings or pairings between each copy. In Appendix \ref{sect:Algorithm_fulldiag} we review the full diagonalisation methodology outlined in Ref. \onlinecite{Kells2014}  which is used to benchmark the commutator algorithm of section \ref{sect:Algorithms}.  In Appendix \ref{sect:AddNum} we outline some additional numerical results and in appendix \ref{sect:subset} we discuss the possibility of using sub-sets of operators to represent the mappings between ground states.

\section{Commutator representations , Quadratic Hamiltonians, and conservation of excitation number}
\label{sect:Opreps_all}

The central aim of this section is to show how the zero-mode solutions of the Hamiltonian commutator $[H,\gamma_{L/R}]=0$, (see for example Refs. \onlinecite{Goldstein2012,Yang2014}), are related to the arguments that establish the Majorana mode stability by proving the universal even-odd degeneracy for all eigenstates of the model.\cite{Kells2014}  In section \ref{sect:Opreps} we show how transitions between energy eigenstates are always eigen-operators of the commutator $\mathcal{H}=[H,\bullet]$, and discuss formally how to give a matrix representation to the commutator.  In section \ref{sect:Majorana_basis} we show how to derive a matrix representation for the commutator using the Majorana position space operators and in \ref{sect:Quadratic}  demonstrate that when $H$ is quadratic, this matrix-representation block diagonalises into sub-matrices in which the conserved quantity is the number of fermions involved in a transition.  In this picture, density-density like interactions will break this exact excitation-number symmetry, coupling different sub-blocks together. This results in a  lifting of degeneracies so that the eigen-opertators of the commutator $\mathcal{H}$ can only take the form of individual transitions $\ket{n}\bra{m}$.  In section \ref{sect:Symmetries} we discuss the symmetry operator responsible for excitation number conservation, showing that in the Majorana basis it counts fermion number and that it can also be understood as a sum of hopping or pairing terms between two related copies of the original Hamiltonian.  In section \ref{sect:SpinKitaev} we outline a spin-representation for the Kitaev chain commutator that is block-diagonal.

\subsection{Operator inner products and representations of the commutator}
\label{sect:Opreps}

\begin{figure}
\includegraphics[width=0.45\textwidth,height=0.4\textwidth]{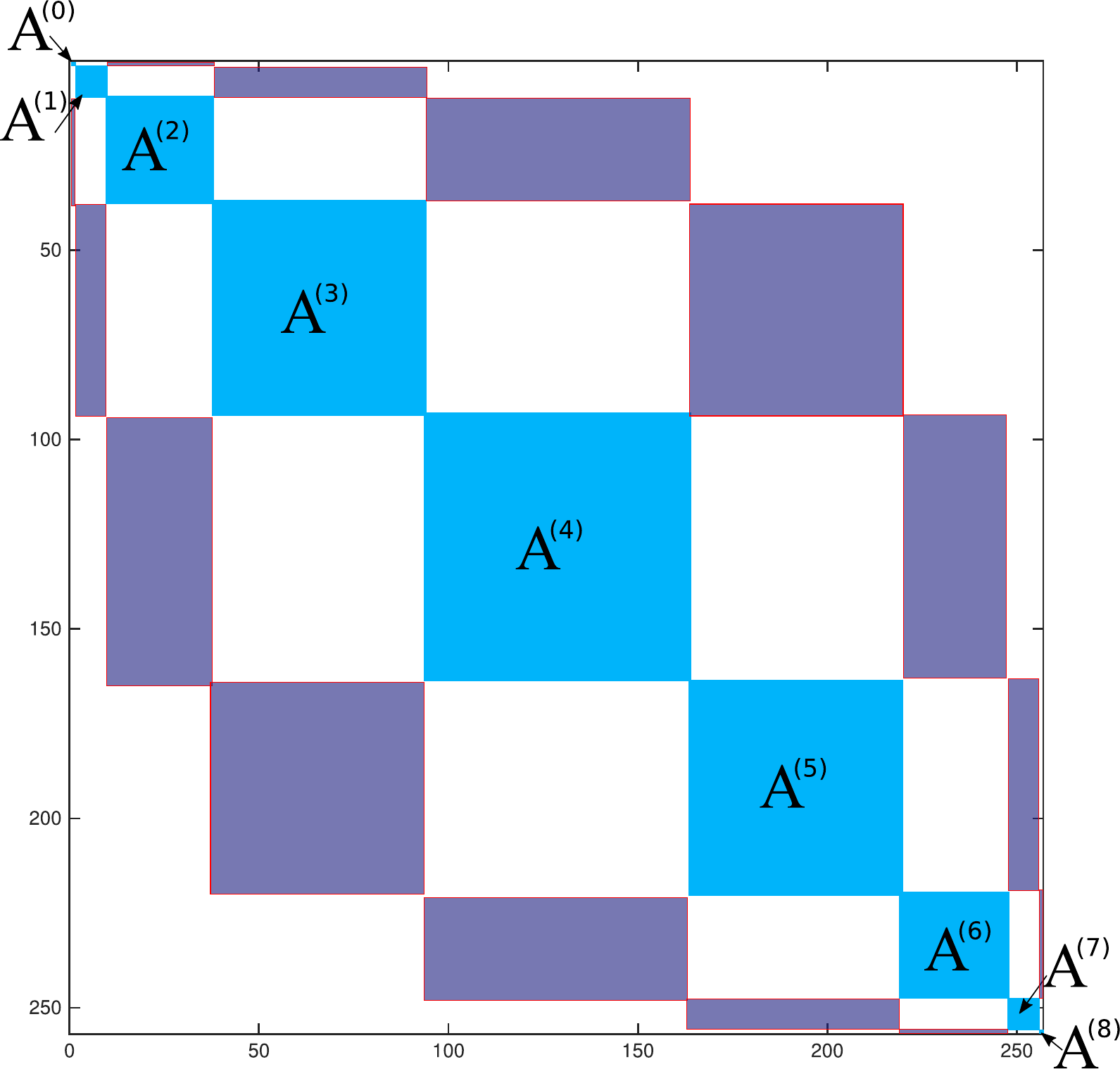}
\caption{The commutator $\mathcal{H}$ for a quadratic fermion model can be decomposed in to blocks $A^{(s)}$ which encode energy transitions involving the same number of fermions $s$. In this example we show the $2N+1$ blocks due to an $N=4$ fermion model. Quartic interacting terms (dark-blue/purple) connect different $A^{(s)}$ blocks, breaking the symmetry responsible for the excitation number conservation.}
 \label{fig:Ablocks}
 \end{figure}

If $\ket{n}$ is an orthonormal basis for a Hilbert space then we can decompose any operator $\hat{X}$ as
\be
\hat{X} = \sum_{nm} \ket{n}\bra{n} \hat{X} \ket{m}\bra{m} =  \sum_{nm} \bra{n} \hat{X} \ket{m}  \ket{n} \bra{m} 
\ee
Using the Hilbert-Schmidt or operator inner-product
\be
\label{eq:IP}
\BraKet{A}{B} = \frac{\Tr (A^\dagger B) } {\sqrt{\Tr(A^\dagger A) \Tr (B^\dagger B)}}
\ee
we have  $\BraKet{\ket{n}\bra{m}} {\hat{X}} = \Tr(\ket{m}\bra{n} \hat{X}) = \bra{n} \hat{X} \ket{m} $ and we 
can rewrite the operator decomposition in a generalised Dirac vector space notation:
\be
\label{eq:OBC}
\Ket{\hat{X} } = \sum_{nm} \BraKet{\ket{n}\bra{m} }{\hat{X}} \Ket{ \ket{n}\bra{m} }
\ee
where the basis states are labeled by operators. 

If we introduce a set of orthonormal vectors  $\Bra{\Psi_i}$ such that $\BraKet{\Psi_i}{\Psi_j}=\delta_{ij}$ we can, by using the cyclic properties of the  trace, also provide a representation for more general operations such as commutators:
\bea
\label{eq:X_super}
\mathcal{X}_{ij} &=&  \Big( \Psi_i  \dashline [X, \bullet] \dashline \Psi_j \Big) \\ 
\non &=& \Big(\Psi_i  \dashline [X, \Psi_j]\Big) = -\Big( [\Psi_i,X^\dagger]  \dashline \Psi_j\big) \\
\non&=& \BraKet{\Psi_i}{X \Psi_j}-\BraKet{\Psi_i X^\dagger}{\Psi_j} \\
\non&=& \Bra{\Psi_i} X^R -X^L \Ket{\Psi_j} 
\eea
where,  to give a matrix representation to the operator $X^L$ we should consider its action to the left . Note that in this case  the conjugate of the $X$ appears to the right of what ever is inside  $\Bra{\Psi_i}$. We can use the above procedure to define the transition Hamiltonian matrix
\bea
\label{eq:H_super}
\mathcal{H}_{ij} &=&  \Big( \Psi_i  \dashline [H, \bullet] \dashline \Psi_j \Big) \\ 
\non&=& \BraKet{\Psi_i}{H \Psi_j}-\BraKet{\Psi_i H^\dagger}{\Psi_j} \\
\non&=& -\Bra{\Psi_i} H^L -H^R \Ket{\Psi_j} 
\eea
In this definition the operators $\Psi$ are labels for vectors within an enlarged Hilbert space where the matrix $\mathcal{H}$ encodes all possible transitions between all eigenstates of the usual Hamiltonian $H$. To see this, note that if $\ket{n}$ are the eigenstates of the Hamiltonian  ($H \ket{n} = E_n \ket{n} $ ) then outer products $  \omega_{nm}  = \ket{n} \bra{m} $ are orthonormal eigen-operators of the commutator with eigenvalues $E_n-E_m$. This can be easily seen from direct calculations
\bea
&& [H, \bullet ] \ket{n} \bra{m}   =   [E_n -E_m]   \ket{n} \bra{m}  \\
&& \Tr ( \ket{l}\braket{k}{n}\bra{m}) = \delta_{nk} \delta_{ml} \non
\eea
or equivalently using the the operator inner-product notation introduced above, note that:
\bea
\mathcal{H}_{nm,kl} &=&-\Bra{\ket{n}\bra{m}} H^L - H^R \Ket{\ket{k}\bra{l}}  \\ \non
&=&  -\Bra{\ket{n}\bra{m}} E_m -  E_k \Ket{\ket{k}\bra{l}}  \\  \non &=&  (E_k - E_m) \delta_{nk}\delta_{ml}
\eea

Clearly then, transition outer products  are eigenstates of the commutator $[H,\bullet]$ and this does not depend on the Hamiltonian being quadratic.  However, for free quadratic systems there will be many eigenstates $\ket{n}\bra{m}$ with the same eigenvalues $(E_n - E_m)$ and a single {\em quasi-particle} excitation with energy $\epsilon = E_n-E_m$ can be understood then as a particular superposition of these degenerate outer-products  $\ket{n}\bra{m}$. In this light we see that the Majorana quasi-particles in Eq. \ref{eq:MBM} are a specific example of this where all $E_n-E_m$ go to zero.  The central result of Ref. \onlinecite{Kells2014} is that this universal even-odd degeneracy remains even in the presence of interactions and therefore there is a well defined and unique definition of the Majorana zero-mode operators throughout the topological region.

\subsection{An operator basis with Majoranas}
\label{sect:Majorana_basis}
In this section we show how by using the fermionic algebra for Majorana fermions we can construct an orthogonal basis for the commutator Hilbert space. For more details readers should consult Ref. \onlinecite{Goldstein2012}. Setting 
\be
\label{eq:Majdef1}
p_n=\gamma_{2n-1} = c^\dagger_n + c_n^\phd , \quad m_n=\gamma_{2n} = i (c^\dagger_n - c_n^\phd )
\ee 
which obey the algebra $\{ \gamma_i ,\gamma_j \}= 2 \delta_{ij}$ and thus $\gamma_i = \gamma_i^\dagger$ and $\gamma_i^2 =I$.  If $N$ is the number of unique fermion modes in our system, using these root operators we can then construct a full set of orthogonal operators:
\bea
\label{eq: MFbasis}
\Gamma^{(0)} : & \quad  & I \\
\Gamma^{(1)} : &  & \gamma_{1},\gamma_{2},\gamma_{3},\dots,\gamma_{2N},\non\\
\Gamma^{(2)} :   &  & i\gamma_{1}\gamma_{2},i\gamma_{1}\gamma_{3},\dots,i\gamma_{2N}\gamma_{2N},\non \\
\Gamma^{(3)} :   &  & -i\gamma_{1}\gamma_{2}\gamma_{3},\dots,-i\gamma_{2N-2}\gamma_{2N-1}\gamma_{2N},\non \\
\vdots \;\;\;\;\;  &  & \vdots \non \\
\Gamma^{(2N)} :  &  & i^{\left(2N\right)N}\gamma_{1}\gamma_{2}\dots\gamma_{2N}\;.\non
 \eea
 We will denote each operator by $\Gamma_{a}$, for $a=1,\dots,2^{2N}$.  For each $a$ one then defines $s$ to be the number of $\gamma$'s in the product $\Gamma_{a}$.  In each of these subsets there are $\binom{2N}{s}$ elements and when we need to refer to a particular element of the subset $s$ we will write $\Gamma_a^{(s)}$.   It may occasionally be convenient  to also use the notation $m_x =\Gamma^{(m)}_x$ and  $p_x =\Gamma^{(p)}_x$ to refer to the specific types of $\Gamma^{(1)}_x$ terms. The phases are chosen so that $\Gamma_a^2 =I$  and  since the product of two $\Gamma_a$'s  gives a third (up to a phase) and $\Tr (\Gamma^{(s)}_a) =0 $ for $s>1$ then we have
\be
\BraKet{\Gamma_a}{\Gamma_b} \equiv \Tr(  \Gamma_a^\dagger \Gamma_b) / 2^{2N} = \delta_{ab} 
\ee 

In Appendix \ref{sect:spinSigma} we show that one can define an orthonormal operator basis with complex fermions  in a similar way to how we defined $\Gamma_a$ above using the algebra of the $\gamma$ terms. We will see that this construction allows one to map the commutator to a normal fermionic system composed of two copies of the original Hamiltonian. We can therefore understand many properties of commutator $[H, \bullet]$ , including the symmetries responsible for its block diagonal structure of non-interacting models, in terms of hoppings and pairings between these two copies.  Furthermore,  complex fermions representations are crucial to understanding how the block structure of the transition matrix $\mathcal{H}$ when written in the $\Gamma$ basis, is related to the single-particle transitions  when the Hamiltonian $H$ is quadratic.  We will discuss this in section \ref{sect:Quadratic}.

\subsection{Quadratic Hamiltonians and Block diagonal Commutators}
\label{sect:Quadratic}

A generic quadratic free-fermion Hamiltonian can be written as
\be
\label{eq:Hquad}
H_{Q} =  \frac{1}{2} \sum_{i,j=1}^N A^{(1)}_{ij} \gamma_i \gamma_j 
\ee
where $A^{(1)}_{ij} =-A^{(1)}_{ji}$ is a pure imaginary number.  For free fermionic systems  it is sufficient to diagonalise the matrix $A^{(1)}$ to be able to write down expressions for all eigenstates of the Hamiltonian.  This is because the eigensolutions of this block represent the single excitations of the free fermionic system:
\be
H_{Q} = \frac{1}{2} \sum_i \epsilon_i (\lambda_i^\dagger \lambda_i - \lambda_i \lambda_i^\dagger) \ee  with
\bea
&&  \lambda^\dagger_i  = \sum_{j=1}^{2N} V^{*}_{ij} \gamma_j,\quad \quad \lambda^\phd_i  = \sum_{j=1}^{2N} V^{}_{ij} \gamma_j.
\eea
The ground state of the system is defined to be the state with zero occupancy of all modes. Higher energy eigenstates are defined by filling some or all of these modes.  

By direct computation using Eqs. \eqref{eq:IP}, \eqref{eq:H_super} and \eqref{eq:Hquad}, the transition Hamiltonian matrix
\be
\mathcal{H}^{\Gamma}_{ab} = \Big( \Gamma_a \dashline [H_{Q}, \bullet] \dashline \Gamma_b \Big)
\ee
can be easily seen to be block diagonal in each of the unique sub-blocks consisting of $a$'s and $b$'s with the same  $s$ (see Fig. \ref{fig:Ablocks} ).   The simplest non-trivial example is the $s=1$ sub-block , which is actually the $2N \times 2N$ adjacency matrix $A^{(1)}$ used to define the Hamiltonian in Eq. \eqref{eq:Hquad} (see Ref. \onlinecite{Goldstein2012}). 

It is helpful to note how $\mathcal{H}$ looks in the complex $\lambda$-fermion basis generated from $I= \lambda^\dagger_n \lambda_n + \lambda^\phd_n \lambda^\dagger_n$, $\lambda_n$, $\lambda_n^\dagger$  and $Z^{(\lambda)}_n= \lambda^\dagger_n \lambda_n - \lambda^\phd_n \lambda^\dagger_n$.   We will refer to generic combinations of these $\lambda$-terms  as $\Lambda_a$ where we again assume a normalisation such that $\BraKet{\Lambda_n}{\Lambda_m} =\delta_{nm}$. In this case, the matrix is diagonal: 
\be
\mathcal{H}_{ab}^\Lambda =  \Bra{\Lambda_a} [H_{Q}, \bullet] \Ket{\Lambda_b} = \delta_{ab} E_a
\ee
where $E_a$ is a weighted sum over energies $\pm \epsilon_n$ for each un-paired $\lambda_n^\dagger$ and $\lambda_n$ occurring in the element $\Lambda_a$:
\be
E_a =\sum_{\lambda_{i}^{\dagger} \in  \Lambda_a} \epsilon_i - \sum_{\lambda_i \in  \Lambda_a} \epsilon_i 
\ee

The states $\Ket{\Lambda_a}$  labelled with $\lambda_n^\dagger, \lambda_n, I_n $ or $Z_n$  can be identified with eigenstates of just one of the $A^{(s)}$ sub-blocks. However if we  use the basis set $\lambda_n^\dagger, \lambda_n, \lambda_n^\dagger \lambda_n $ and $\lambda_n^\phd \lambda_n^\dagger$ this is not the case.  To illustrate this we consider the example of the state $ \Ket{\Lambda_a} = \Ket{ \lambda^\dagger_1 , \lambda_2 ,   Z_3, I_4 }$ which is a transition energy eigenstate of the $A^{(4)}$ block with an energy $\epsilon_1-\epsilon_2$ .  Importantly this state can be written as a superposition of two states 
\bea
 \Ket{\Lambda_a} &=&\Ket{ \lambda^\dagger_1 , \lambda_2 ,   Z_3,  I_4 } \\
 &=& \Ket{ \lambda^\dagger_1 , \lambda_2 ,   \lambda_3^\dagger \lambda_3- \lambda_3^\phd \lambda^\dagger_3,  I_4 } \non \\ 
  &=& \frac{1}{\sqrt{2}} (\Ket{ \lambda^\dagger_1 , \lambda_2 ,   \lambda_3^\dagger \lambda_3,  I_4 } - \Ket{ \lambda^\dagger_1 , \lambda_2 ,   \lambda_3^\phd \lambda^\dagger_3,  I_4 }) \non
 \eea
which both have support on the $A^{(2)}$ and $A^{(4)}$ sectors. When brought together with a negative sign like this, the parts of the wave-function in the $A^{(2)}$ sector cancel and we are left with only the part of the state in the $A^{(4)}$ sub-block.  On the other hand suppose we bring these states together with a positive sign. In this case we have
\bea
 \Ket{\Lambda_a}  &=& \frac{1}{\sqrt{2}} (\Ket{ \lambda^\dagger_1 , \lambda_2 ,   \lambda_3^\dagger \lambda_3,  I_4 } + \Ket{ \lambda^\dagger_1 , \lambda_2 ,   \lambda_3^\phd \lambda^\dagger_3,  I_4 }) \non \\
  &=& \Ket{ \lambda^\dagger_1 , \lambda_2 ,   \lambda_3^\dagger \lambda_3+ \lambda_3^\phd \lambda^\dagger_3,  I_4} \non \\
  &=&\Ket{ \lambda^\dagger_1 , \lambda_2 ,   I_3,  I_4 } 
 \eea
which is entirely supported by the $A^{(2)}$ block.

\vspace{2mm}
In section \ref{sect:Opreps} we noted that the eigen-operators of $\mathcal{H}$ are the outer-products $\ket{n}\bra{m}$ and that the eigenvalues are $E_m-E_n$.  However, if the fermionic system is quadratic, we can also solve in each of the $A^{(s)}$ sub-blocks separately and we know that we can interpret the solutions of the single-particle sub-block $A^{(1)}$ as operators which act across the full Hilbert space. To see how these two pictures are related we need to understand exactly what is happening in all sub-blocks.

The solutions of the $A^{(1)}$ sub-block represent all possible transitions between states which differ by the occupancy of a single $\lambda_i$ or $\lambda_i^\dagger$ mode.  The excitation energies $\pm \epsilon_i  $ therefore correspond to the energy difference $\pm (E_n -E_m)$ between any two such states.  Importantly, because this sub-block is spanned only by elements $\Ket{\gamma_1, I_2,I_3,...,I_N}$ , $\Ket{ I_1,\gamma_2,I_3, ... ,I_N}$ etc. that have only single the entries of $\gamma$ combined with $I$'s on all other sites, we write our eigensoluitions of this block as (e.g $\Ket{\lambda_1^\dagger , I_2, I_3 ,....,I_N}$  ,  $\Ket{I_1 , \lambda_2, I_3 ,....,I_N}$ etc. with energies  $\epsilon_1$, $-\epsilon_2$  resp. ) in a similar fashion.  

In the more general cases of odd valued $s$, we see that the transition energies of the state $\Ket{\Lambda_a}$ only depend on the number of  unpaired $\lambda_i^\dagger$'s and $\lambda_i$'s and therefore states like $\Ket{\lambda_1 ,Z_2,I ... I}$ ,$\Ket{\lambda_1 ,I ,Z_3 ... I} $, $\Ket{\lambda_1 ,Z_2,....,Z_N}$, etc.   all have energy $-\epsilon_1$. They are therefore  degenerate with the $-\epsilon_1$ state which is contained fully within the $A^{(1)}$ sector.  In each sector there are  $\binom{N-1}{(s-1)/2} $ states with the same energy as the single particle transition $\epsilon_1$ and therefore associated with a unique excitation $\lambda_1$, we have total of %
\be
\sum_{s\in odd} \binom{N-1}{ \frac{s-1}{2}} =2^{N-1}
\ee
transitions. 

Recall now our original eigenbasis states $\ket{n}$. In the eigenbasis provided by single particle $\lambda_i$ fermion states we may associate each state $n$ with a Fock representation such that
\be
\ket{n} = \ket{ n_1, n_2, n_3, ........,n_N }
\ee
is given by the binary occupation number of the fermionic levels $\lambda_i$. Using  \eqref{eq:sigmarules} this means for example that, we can write an individual outer-product transition 
\bea
\Ket{\ket{10...1}\bra{00...1}} =  \Ket{ \ket{1}\bra{0}} \Ket{ \ket{0}\bra{0}} .... \Ket{ \ket{1}\bra{1}} , \non
\eea
as
\bea
\Ket{ \lambda^\dagger_1, \lambda_2^\phd \lambda_2^\dagger , ....,,\lambda_N^\dagger \lambda_N^\phd }
\eea
  Again we stress that although this transition is an eigenstate with the energy $\epsilon_1$, it is not the same as the state $\Ket{\lambda_1^\dagger, I_2, ...... I_N}$ which represents the actual free $\lambda_1^\dagger$ transition that we calculate by diagonalising the $A^{(1)}$ block which encodes the BdG/Majorana matrix forms.

\subsection{Symmetries and hopping/pairing}
\label{sect:Symmetries}
 The Hamiltonians we study in this paper all conserve fermionic parity. That is, each term appearing in the Hamiltonian, is constructed from a product of an even number of fermionic terms. This means that the transition Hamiltonian $\mathcal{H}=[H,\bullet]$ can always be decomposed into two sectors: an odd sector and an even sector.  This parity conservation is due to a symmetry 
\be
\mathcal{P}=\prod_j p^L_j m^L_j p^R_j m^R_j = P^L P^R
\ee
where $P^L$ is the parity operator for $H^L$ and $P^R$ is the parity operator for $H^R$. 

When the system is strictly quadratic, each of these excitation-parity sectors can be further decomposed into smaller blocks  which are spanned by basis states $\Ket{\Gamma^{(s)}}$. This block diagonal structure exists because of a symmetry of the  commutator $\mathcal{H}$ in which the conserved quantity is the total number of particle excitations and de-excitations e.g. the eigensolutions of the $A^{(1)}$ block represent all possible single particle transitions , the solutions of the $A^{(2)}$  all double transitions  etc.).  The operator responsible should commute with the $\mathcal{H}$  and, if well chosen, count the number of unique $\gamma$ terms in each basis state $\Ket{\Gamma}$.  Working under the assumption that left and right acting operators are mutually fermionic, the symmetry operator $\mathcal{N}$ then in this case is 
\bea
\mathcal{N} &=&   \sum_{j=1}^N I+ \frac{ i}{2}  (p_j^L p_j^R -  m_j^L m_j^R) \\ &=& \sum_{j=1}^N I-  \frac{i}{2}( c_{j}^{\dagger L}c_{j}^{\dagger R}+ c_j^{L}c _j^{R} )\non.
\eea
We can easily see that this operator does indeed count the number of excitations. For example using the spin representation outlined in section \ref{sect:spinGamma}  each term in the above summation locally looks like
\be
\frac{2 I \otimes I - I\otimes \sigma^z+ \sigma^z \otimes I}{2} = \left[ \begin{array}{cccc} 0 & 0& 0&0 \\ 
 0 & 1 & 0&0 \\ 
 0 & 0 & 1&0  \\ 
 0 & 0 & 0&2  \end{array} \right] 
\ee
where the basis states are given by Eq. \eqref{eq:Gamma_basis}.
 
\subsection{A block-diagonal spin representation for the Kitaev chain commutator $[H,\bullet]$}
\label{sect:SpinKitaev}

\begin{figure}
\includegraphics[width=0.36\textwidth,height=0.47\textwidth]{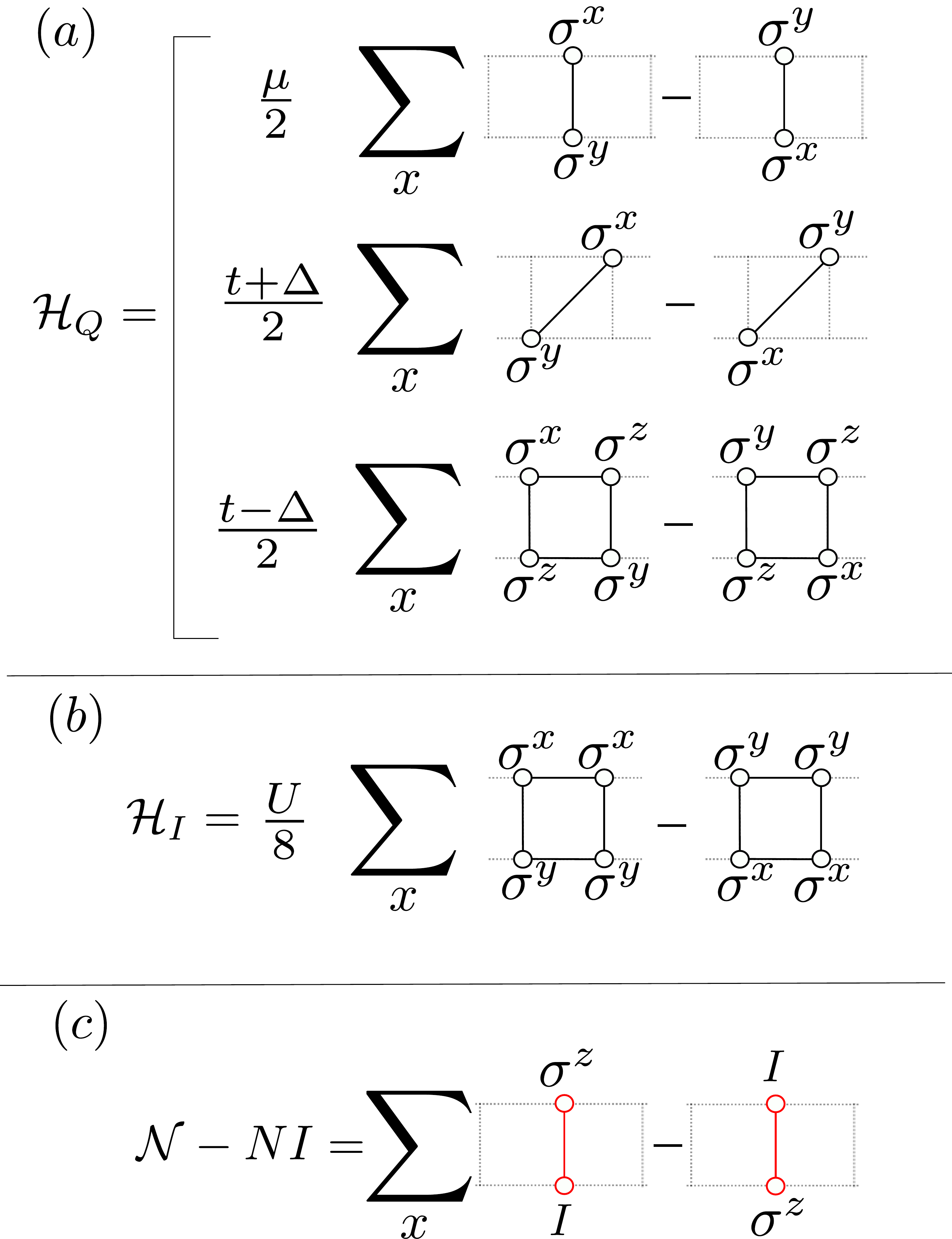}
\caption{(a) Graphical representation of the block diagonal $\mathcal{H}_Q$. (b) Graphical representation of the quartic term $\mathcal{H}_I$. (c) The symmetry operator responsible excitation number conservation of $\mathcal{H}_Q$. }
 \label{fig:model}
 \end{figure}

Using the basis $\Gamma$-basis  outlined in  Appendix \ref{sect:spinGamma}  we can write the commutator $[H,\bullet]$ for the 1-d $p$-wave chain (see Fig. \ref{fig:model} and Appendix \ref{sect:Kitaevchain}) as:
\bea
\label{eq:HQcommutator}
\mathcal{H}_{Q} &=&  \frac{u_l}{2} \sum_j (\sigma^x_{2j-1} \sigma^y_{2j} - \sigma^y_{2j-1} \sigma^x_{2j}) \\  \non
&+&   \frac{t+\Delta}{2}  \sum_j (\sigma^y_{2 j} \sigma^x_{2j+1} - \sigma^x_{2 j} \sigma^y_{2j+1} )\\ \non
&+& \frac{t-\Delta}{2} \sum_j \sigma^z_{2j} \sigma^z_{2j+1} (\sigma^x_{2j-1} \sigma^y_{2 j+2} -\sigma^y_{2j-1} \sigma^x_{2 j+2} )
\eea
and 
\be
\mathcal{H}_I = \frac{U}{8} \sum_j (\sigma^x_{2j-1} \sigma^y_{2j} \sigma^x_{2j+1}\sigma^y_{2 j+2} -\sigma^y_{2j-1} \sigma^x_{2j} \sigma^y_{2j+1}\sigma^x_{2 j+2})
\label{eq:HIcommutator}
\ee

The matrix  $\mathcal{H}_Q = [H,\bullet]$ decomposes into blocks that count the overall excitation number, see section \ref{sect:Symmetries}. Note that the spin-representation used above is by no means unique. We use this one because here  the symmetry $\mathcal{N}$ is diagonal and there is a direct relationship between the binary indexing of the basis elements and and the precise form of $\Gamma_a$. For example, state $\Ket{m_1}$ has a binary index $100 ... 000$,  $\Ket{p_1}$ has a binary index $010 ... 000$, and $\Ket{Z_N} $ has the index $000 ... 0011$.  This makes it easy to interpret the meaning of the eigenstates of $\mathcal{H}$.

In section \ref{sect:Quadratic} we discussed how the actual eigenstates of the commutator $\mathcal{H}_Q$, which encode all transitions in our free system, are related to solutions obtained by diagonalising each block separately. Indeed, as is well understood, for a quadratic system we only need to focus on the $A^{(1)}$ sub-block which is just a representation of the original adjacent matrix used to define the full Hamiltonian.  However the actual transitions from one state to another $\ket{n}\bra{m}$ are superpositions of degenerate excitations taken from all of the blocks $A^{(n)}$ with the corresponding parity. In contrast, the actual quasi-particle operators themselves (and combinations of them) are contained inside the blocks.  While the introduction of an interacting term will break this block diagonal structure, see FIG. \ref{fig:Ablocks}, the above observation shows us that the many-body Majorana operators , which by definition are  superpositions of odd-parity zero energy transitions,  should be predominantly supported within the $A^{(1)}$ sector and therefore they are the odd-parity zero-energy modes that minimise their $\mathcal{N}$ expectation values. This notion forms the basis for the algorithm that we define in the next section.

\section{Algorithms and numerics}
\label{sect:Algorithms}

\subsection{Algorithms for computing zero-energy Majorana modes in the presence of local interactions}
\label{sect:CM}

In Ref. \onlinecite{Kells2014} it was demonstrated that , in the $L \gg \xi$ limit, there was a well defined notion of the Majorana quasi-particle even in the presence of strong interactions.  The stability of the Majorana to strong interactions follows from the fact that the degeneracy between all even-odd pairs remains to an order of perturbation theory that scales with the length of the system.   Using the definition of Eq. \ref{eq:MBM}  the position space many-body Majorana wave functions can then be calculated using the Trace (or Hilbert-Schmidt) inner-product. We have
\bea
\label{eq:GammaLR}
u^{(n)}_R(\vec{x})&=&\Tr ( \Gamma^{(n)}_x \times \gamma_R)/ 2^{N} \equiv \BraKet{ \Gamma^{(n)}_x  }{\gamma_R} \\ \non
u^{(n)}_L(\vec{x})&=&\Tr ( \Gamma^{(n)}_x \times \gamma_L) /2^{N} \equiv \BraKet{ \Gamma_x^{(n)}   }{\gamma_L} .
\eea

The full diagonalisation method (FD)  is reviewed  again in the Appendix \ref{sect:Algorithm_fulldiag}.  This method is accurate but limited to small system sizes. This is because, although the Hamiltonian is sparse, the eigenvectors are not and we need all of them.  In order to go to larger system sizes where we can probe systems with longer coherence lengths we need another method.  The approach we use is to focus on the $\Gamma$-representations of the commutator $\mathcal{H}=[H,\bullet]$. We call this procedure the commutator method (CM).

Like with the FD method, the key index in the CM algorithm is the number of position space sites $N$. The matrix representation scales as $2^{2 N}$ where  each $A^{(s)}$ block is spanned by $\binom{2N}{s}$ basis elements.  In order to proceed we introduce physically motivated cut-offs for the  number of basis elements that are needed to accurately represent the Majorana quasi-particle.  The first  cut-off $N_s$ represents the maximum number of blocks $A^{(s)}$ that participate in the calculation. Thus an excitation number cut-off of $N_s=3$ would only allow elements of $A^{(1)}$ and $A^{(3)}$, and a cut-off of $N_s=9$ will only allow elements from blocks $A^{(1)},A^{(3)},A^{(5)}, A^{(7)}$ and $A^{(9)}$.   The second cut-off is one where we only allow basis elements which can be reached by $N_d$ operations of the Hamiltonian on full set of of single-excitation basis elements that span $A^{(1)}$. This cut-off is in some ways similar to to $N_s$ but allows us to kick out basis elements used inside each of the $A^{(s)}$ sectors that are not in anyway close to the single-particle block.  The algorithm is therefore in some way perturbative and we do not expect it to be accurate for large interaction strengths.  However, the truncation errors introduced can easily be controlled by  demanding that results converge for sufficiently high cut-off values  $N_s$ and $N_d$.

While the cut-offs above allow us to fit the problem on a computer, another challenge is actually finding the particular zero-valued eigenstates of reduced operator $\mathcal{H}_{\text{red}}$ which correspond to the Majorana quasi-particles.  As we mentioned in the previous section, the general method is to find the zero-valued solutions that minimise their expectation value of $\mathcal{N}$. While this is in principle straightforward, there are convergence issues related to the fact that the MBM zero-modes are sitting amongst many other zero and near-zero eigenstates of $\mathcal{H}_{\text{red}}$. To overcome this we have found several robust methods which are in general agreement across a range of parameters.

The first and most straightforward method is to employ a Lanczos diagonalisation where the initial vector is chosen to be one of the two non-interacting modes. Using this method, with multiple restarts, we can rapidly find the eigenstates with the correct properties, provided we have chosen $N_s$ and $N_d$ so that sufficient support is given to the many-particle structure of the mode. For small system sizes where it can be checked, the method gives results that are identical to that of the FD method, see Figure \ref{fig:N10exact}.  We have also checked this procedure against two others. The first of these to evolve in imaginary time using sparse matrix recursive implementations of $\exp (- \mathcal{H}^2 \tau)$ from an initial vector which  is again one of the Majorana modes in the non-interacting regime. Given sufficiently large $\tau$ and high cut-offs off both $N_s$ and $N_d$ we typically see the convergence of energy eigenvalues to a value close to zero and an expectation value for $\mathcal{N}$ between $1$ and $2$.  Recall that there are no other near-zero energy modes with this value of expectation value and therefore we know that if these two criteria are met that the results are an accurate representation of the true-many-body Majorana.  The second method is to again use a Krylov subspace technique (in this case Arnoldi ) to find as many near-zero eigenstates of $\mathcal{H}_{\text{red}}$ as possible.  We then  search for the superposition of these states which minimises its expectation value of $\mathcal{N}$.

\section{Numerical results}
\label{sect:Numbers}
\begin{figure}
\includegraphics[width=0.45\textwidth,height=0.35\textwidth]{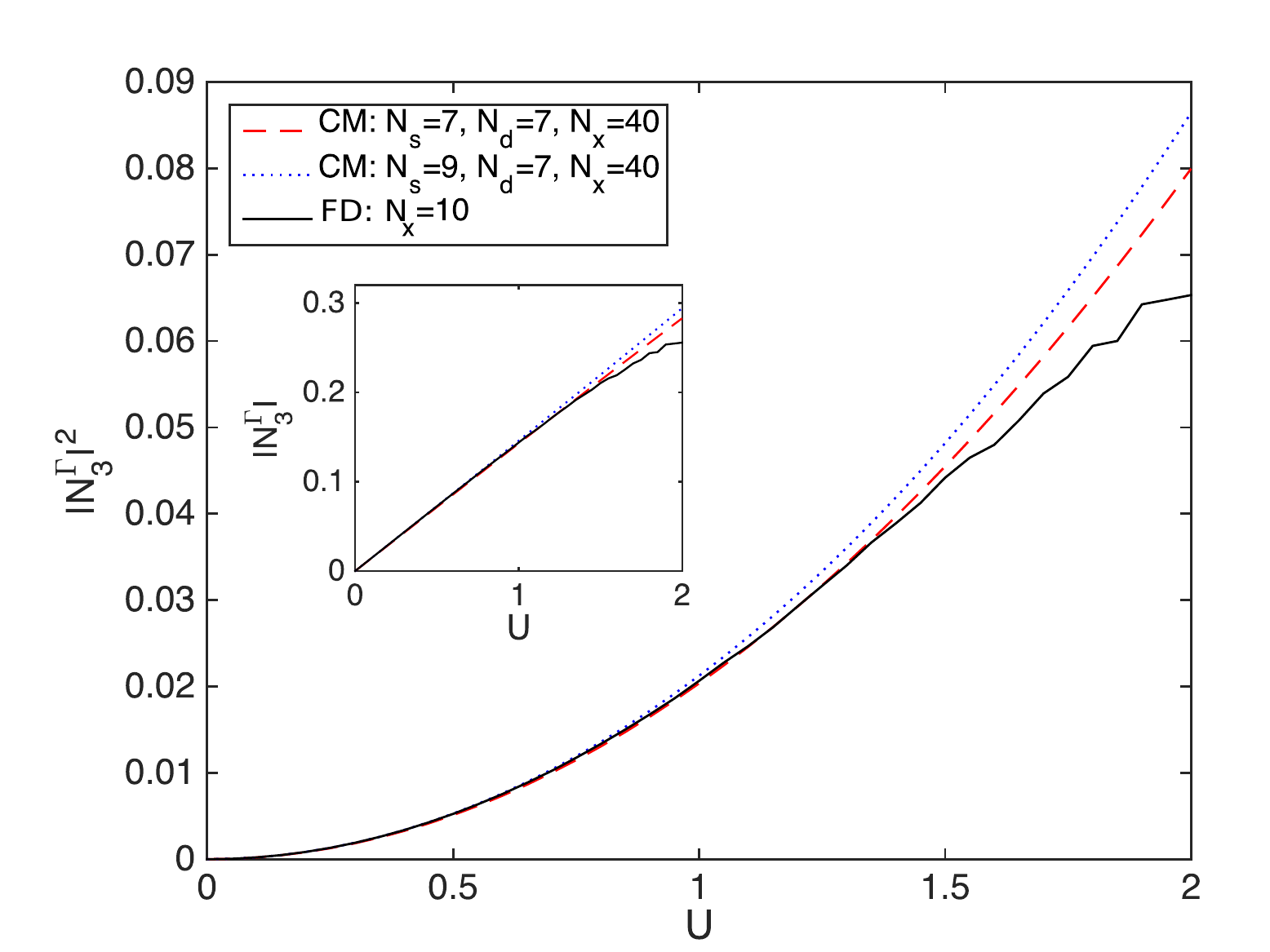}
\caption{In the figure we show the behaviour of $ |N^\Gamma_3|^2$ for a system with $\Delta=0.9$, $\mu=1.5$ and $t=1$ .  For $U=0$ the Majorana mode has a coherence length of $\xi \approx 1.11$ and thus for small values of $U$ permits us to use the full diagonalization method (FD) of  Ref. \onlinecite{Kells2014}. For this purpose we use a system size of $N_x=10$. We compare this with results using the commutator method (CM) with $(N_s,N_d)=(7,7)$ and $(7,9)$  for a system of size $N_x=40$.  In the inset we plot the $ |N^\Gamma_3|$ to emphasise the linear dependence on $U$.  }
 \label{fig:N10exact}
 \end{figure}
\begin{figure}
\includegraphics[width=0.45\textwidth,height=0.35\textwidth]{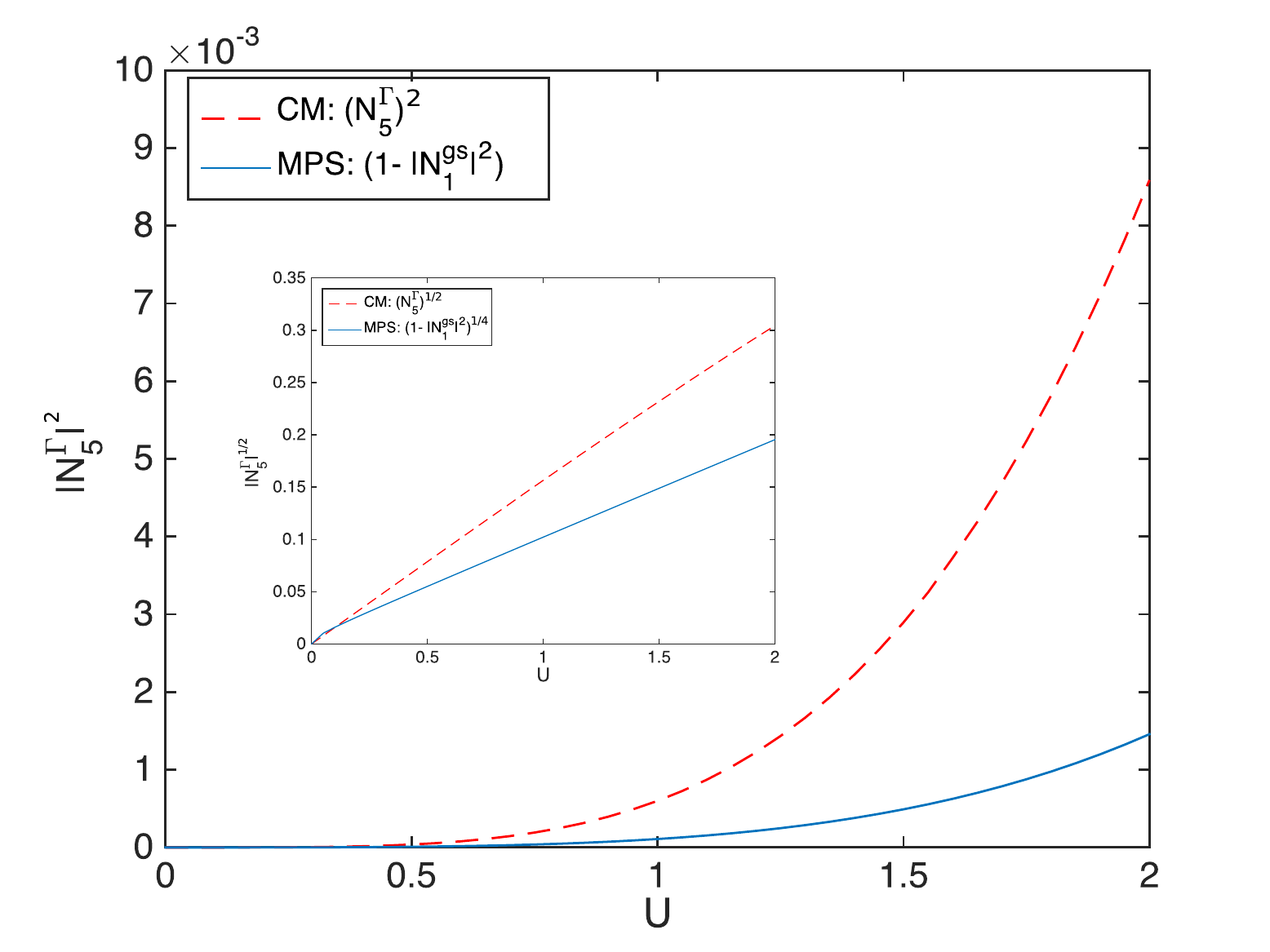}
\caption{In the figure we plot $|N^{\Gamma_5}(U)|^2$ and $1-|N^{gs}_1(U)|^2$ for a system of $N=40$ with $\Delta=0.9$, $\mu=1.5$ and $t=1$.  The inset shows that for these parameters  $(1-|N^{gs}_1(U)|^2)$  and $|N^\Gamma_5|^2$ grow as the 4th power of $U$. The quartic dependence of $1-|N^{gs}_1(U)|^2 $ would seem to indicate that single particle contributions should dominate even at higher interaction strengths (Note the small scale on the Y-axis of the main figure). However, as we discuss in section \ref{sect:correlators} the linear correlators are not a reliable indicator of the N-particle content of the quasi-particle itself. Indeed we note that the $|N^\Gamma_5|^2$  shown here is far smaller that  $|N^\Gamma_3|^2$ shown in Figure \ref{fig:N10exact} for the same system parameters.  }
 \label{fig:MPS_small}
 \end{figure}
\begin{figure}
\includegraphics[width=0.45\textwidth,height=0.35\textwidth]{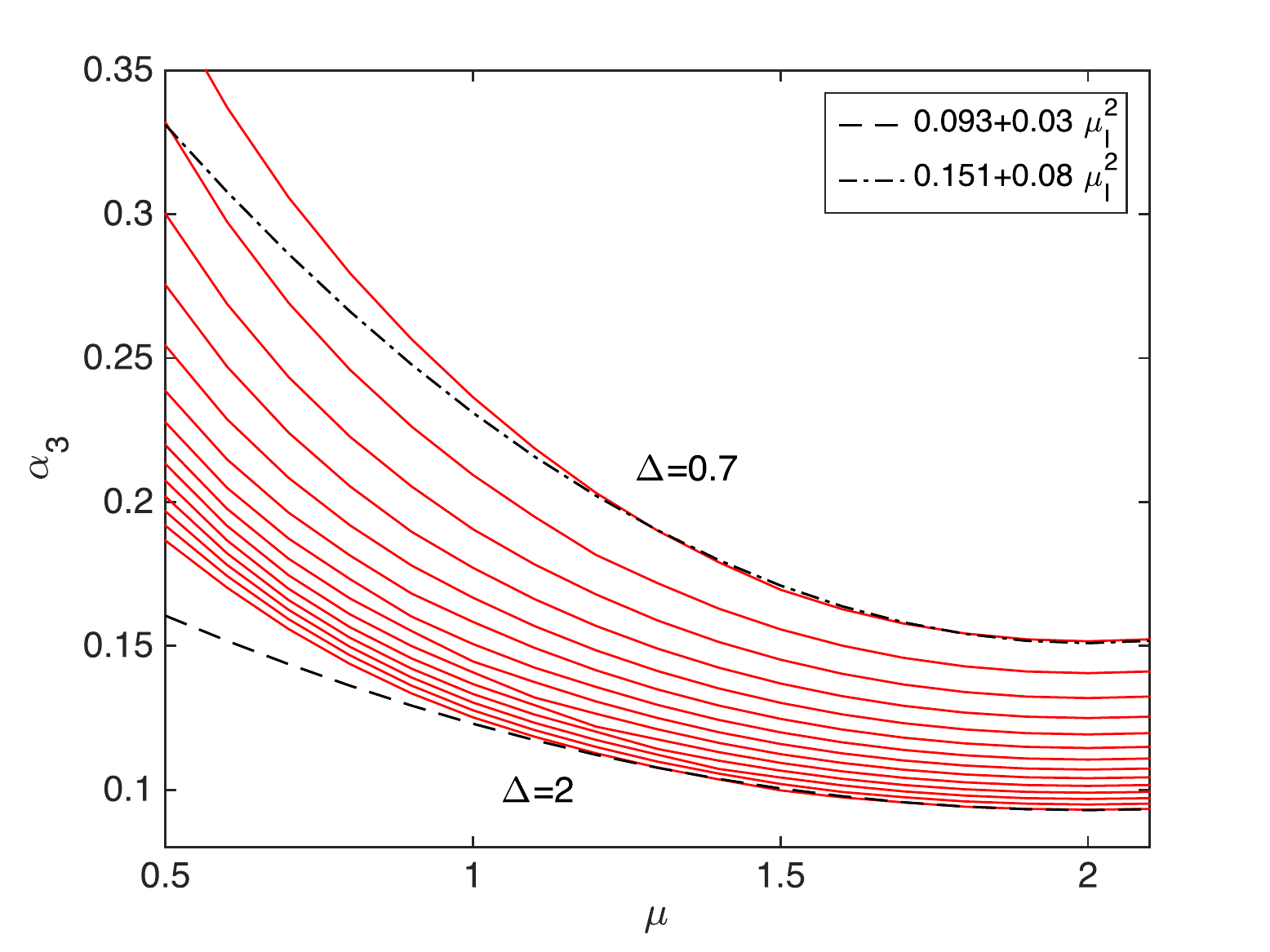}
\caption{In the figure we plot $\alpha_3$, the rate of slope of the $|N_3^\Gamma(U)|$  as a function of $\mu$. Our results show best convergence for small coherence lengths and values of the $\mu$ that are far from the bottom of the band. We see a clear quadratic dependence centered around  $\mu=2t$. The rate of growth of this term is therefore a minimum when we can linearise our dispersion. Fits of quadratic curves  that show good overlap with the numerical data are also given.  A system size of $N_x=50$, with cutoffs $(N_s,N_d)=(9,9)$ was used for this plot. }
 \label{fig:alpha3_muplot}
 \end{figure}
\begin{figure}
\includegraphics[width=0.45\textwidth,height=0.35\textwidth]{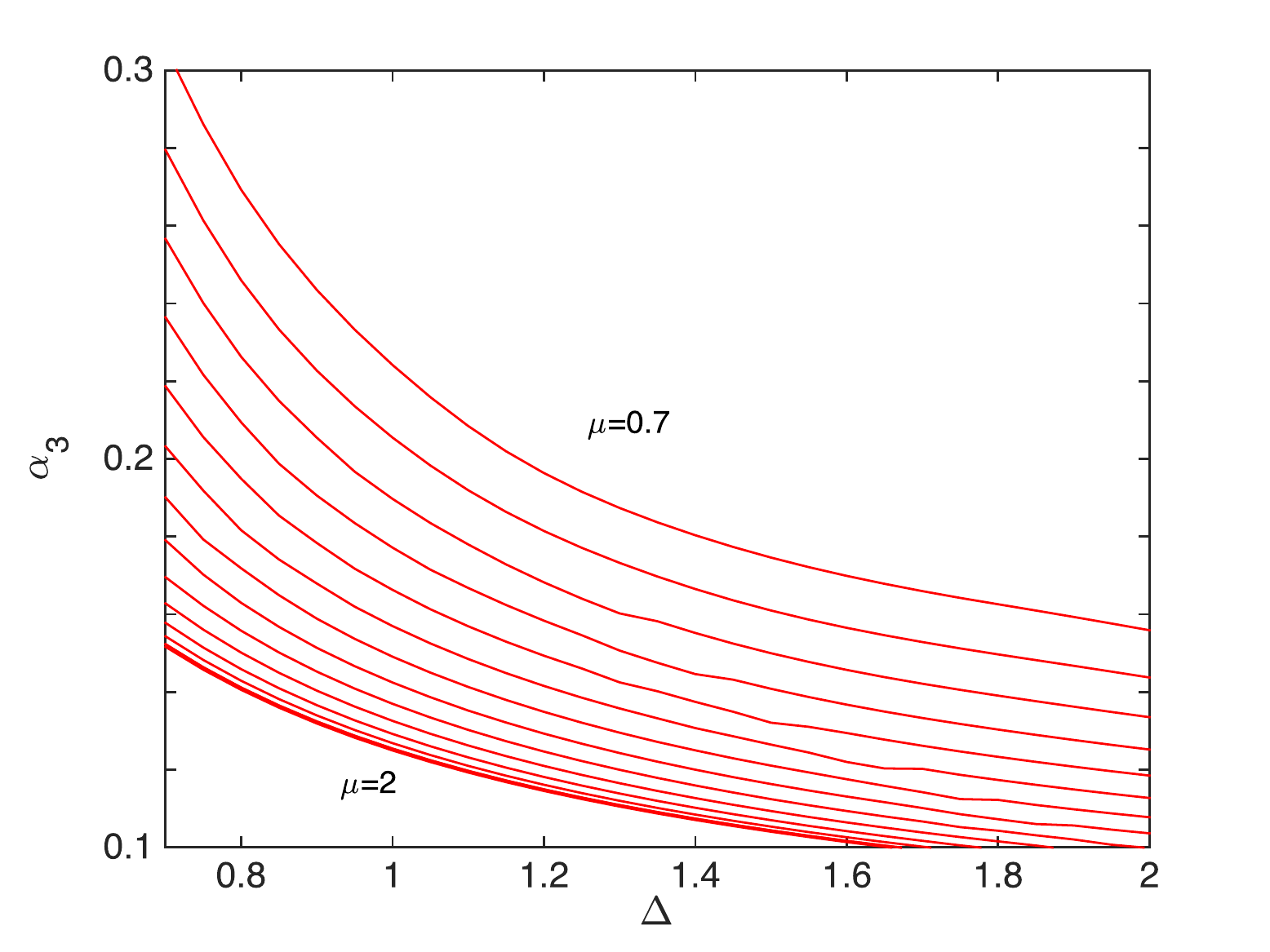}
\caption{In the figure we plot the rate $\alpha_3$ as a function of $\Delta$.  The rate of growth clearly increases as $\Delta$ decreases.  The N-particle content of the many-body Majorana is therefore clearly grows with the increased coherence length associated with a smaller  superconducting gap.  A system size of $N_x=50$, with cutoffs $(N_s,N_d)=(9,9)$ was used for this plot. }
 \label{fig:alpha3_deltaplot}
 \end{figure}

Our main numerical focus is on the N-particle participation ratios of the Majorana quasi-particle:
\be
|N^\Gamma_n|^2  = \int  |u_{L/R}^{(n)} (\vec{x})|^2 d\vec{x}
\ee
which have the property that $\sum |N^\Gamma_n|^2 =1$.   Our main findings, summed up in Figure \ref{fig:N10exact} and Figure \ref{fig:MPS_small}, are that 
\bea
|N^\Gamma_3( U)| &\propto&  \alpha_3 U   \\ \non
|N^\Gamma_5(U)| &\propto &  \alpha_5 U^{2} \\ \non
|N^\Gamma_7(U)| &\propto &  \alpha_7 U^{3} \\
\vdots \non
\eea
for odd values of $n$ greater than $1$.  

The results above follow naturally from the block-diagonal structure of the non-interacting commutator, and the assumption that the MBM's are the zero-modes that minimise the excitation number.  The non-interacting Majorana is contained fully within the $A^{(1)}$ block and it continuously deforms away from this idealised state as we turn on the interaction.  It is important to stress here that the block diagonal structure is not unique to this model and a similar structure exists for any quadratic model modified by a quartic interacting term  We should therefore expect similar scaling behaviour of the multi-particle content in other models of topological superconductivity that permit strong MBM zero-modes.

This intuitive picture also suggests that a perturbative approach could be used to calculate the $\alpha$ coefficients. This has however proved problematic, although work on this is on-going. The main problem is that, in the commutator picture, the zero-energy MBM's sit amongst many {\it almost} zero-energy states, and this causes infinities to appear in the perturbative expansion. It is this band of states that also leads to slow convergence-rates for the purely numerical approaches. In those cases however we have been able to overcome the problem for limited regions of the parameter space.  

The linear $U$ dependence shows that the 3-particle contributions to the many-body quasi-particle itself are relevant even for small interactions strengths. In Figures \ref{fig:alpha3_muplot} and \ref{fig:alpha3_deltaplot} we plot the dependence of the numerically calculated $\alpha_3$ growth rate, which dictates how fast the higher 3-particle contributions to the Majorana quasi-particle grow as we increase the interaction strength $U$.  We see from Figure \ref{fig:alpha3_muplot} that there  is a clear parabolic dependence centered around $\mu=2t$. This value of the chemical potential corresponds to the half-filled band, where, in the lattice model, the energy-momentum dispersion is exactly linear.  This shows that the growth of higher $N$ contributions is sensitive to the precise nature of the underlying dispersion. Figure \ref{fig:alpha3_deltaplot} also shows the clear reduction of the $\alpha_3$ growth rate as the superconducting gap is made larger.  The same general trends are also seen for the $\alpha_5^{1/2}$ parameter, see appendix \ref{sect:AddNum}. 

In Figure \ref{fig:MPS_small}  we also compare an example $|N^{\Gamma}_5|^2$ with the measure $1-|N_\text{gs}|^2$ where
\be
|N^{\text{gs}}|^2  = \int  |O_{L/R}^{\text{gs}}(x)|^2 dx
\ee
and where $O$ is the single-particle operator expansion of the ground state outerproducts such that
\bea
\label{eq:GLR}
|O^{\text{gs}}_{R/L}(x)| &=& | \bra{0_e} c^\dagger_x \pm c_x \ket{0}_o| \\ \non
&=&|\Tr (  c^\dagger_x \pm c_x  \times \ket{0_o}\bra{0_e}  )| 
\eea
To calculate the ground state correlators we use a customised MPS algorithm similar to that outlined in Ref. \onlinecite{Verstraete2008}.  In general agreement with Ref. \onlinecite{Stoudenmire2011}, which calculated the same measure for the related proximity-coupled semiconductor model, we see that  $|N^{\text{gs}}|$ tends to stay very close to unity even at higher interaction strengths. On the surface this would seem to imply a much lower multi-particle contributions than that predicted using the FD and CM methods above. In section \ref{sect:correlators} we show that while $ |O^{\text{gs}}(x) | $ may be a valuable measure for understanding the general position space spread of the Majorana operators,  it is not a reliable indicator of the N-particle participation rates in the quasi-particles themselves. 

Before moving on we note that the trace expression in Eq. \eqref{eq:GLR} does not contain the same factor $1/2^{N}$ as the expressions in Eq. \eqref{eq:GammaLR}.  Then \eqref{eq:GLR} actually represents the single particle operator expansion of the ground-state outer product  $\ket{0_o}\bra{0_e}$ that has been multiplied by a factor of $2^{N}$. We will see in the next section, that only in the case of a non-interacting system does this magnified single-particle expansion correspond to the structure of the quasi-particle.

\section{Measuring the many-body Majorana structure with DMRG and MPS}
\label{sect:correlators}

When the interactions are included, the linear form  \eqref{eq:Majdef} is no longer sufficient to fully describe the Majorana zero-mode. The operators, which are Hermitian, particle-hole symmetric and have odd parity (i.e. the switch the parity of the underlying states) can be expanded in multinomials of terms with odd fermionic parity, see Eq. \ref{eq:Majdef2}.   

The Majorana operator takes us from one ground state to the other $ | \bra{0_e} \gamma_{L/R} \ket{0_o}| =1 $.  For the non-interacting system one way to read-off the full Majorana operator structure is  to examine the correlations   $ \phantom|_e \bra{0} \Gamma^{(1)}_a \ket{0}_o$  where  the ground states $\ket{0}_{e/o}$ are obtained from variational techniques such as DMRG or MPS, and the $\Gamma_a$ operators are restricted to the $2N$ single particle operators $\gamma_i$ .  In the non-interacting limit, the values obtained from the analysis of these single-particle cross-correlators are enough to fully determine the structure of the Majorana operators. The calculation on the left hand side of the system is as follows:
\bea
O_L(x') &=&  \bra{0_e} \Gamma^{(m)}_{x'} \ket{0_o} = \bra{0_e} m_{x'}  \gamma_L \ket{0_o}  \non \\
 &=&  \sum_x u_L(x)  \bra{0_e}m_{x'}m_{x} \ket{0_e} =   u_L(x') \non
\eea
The last equality follows because, for arbitrary normalised states $\ket{\psi}$,   $m_{x_1}\ket{\psi}$ and $m_{x_2} \ket{\psi}$ are orthonormal ( i.e.  $\bra{\psi} m_{x_1} m_{x_2} \ket{\psi} = \delta_{x_1,x_2}$ see below).  The calculation shows that, by examining the cross-correlators $ \bra{0_e}  \Gamma^{(1)}_{x'} \ket{0_o}$ we can learn the form of the non-interacting Majorana operators.  Of course this approach is not really necessary for the non-interacting system, as we can also work out the free fermion excitations from the single transition BdG/Majorana representations.

It is often assumed that because this method  works in the non-interacting regime, it should work equally well  in the presence of interactions.  We now show that this assumption is wrong and that the cross-correlations cannot be used to resolve the precise form of the many-body Majorana (\ref{eq:Majdef2}). We can understand this on a basic level by just noting that the ground-state outer-products are not the same as the Majorana expansions formed using all the eigenstates of the system, see Ref. \onlinecite{Kells2014}. However, it is also illustrative to observe where the simple calculation presented above breaks down when interactions are present.  We will see that in this case, because the Majorana operators will now contain contributions from operators like $ \Gamma^{(3)}_x, \Gamma^{(5)}_x, $ etc. (see Eq. \eqref{eq:Majdef2}) , and because a more complicated set of orthonogonality relations exist  between generic states  $ \Gamma^{(n)}_x \ket{\psi}$ and $ \Gamma^{(m)}_{x'} \ket{\psi} $ , that 
\be
\bra{0_e}  \Gamma^{(n)}_x  \ket{0_o}  \ne  u^{(n)}(x) .
\ee

To see these general orthonormal conditions lets suppose we have an operator $W$ that has been constructed from two odd-number sequences of $\gamma$ operators $W= \Gamma^{(l)} \Gamma^{(m)}  $.  Now consider when, for arbitrary real states $\ket{\psi}$, does the correlation $\bra{\psi} W \ket{\psi}$ vanish. The constituent operators $\gamma$ are unitary operators and they therefore take any position space {\em basis} element to an orthogonal basis element with opposite occupancy on the sites where the operator W has acted:$\braket{n_x}{n'_x}=\bra{n_x} W \ket{n_x} =0 $.  Here the $n_x$ are binary number sequences indicating the occupancy on each position space site. 

As the p-wave Hamiltonian can always be made real, \cite{Kitaev2001} all eigenstates are also real.  For any operator $W$ we can therefore always decompose any eigenstate $\ket{\psi}$ as
\be
\ket{\psi} = \sum_n (a_n  + \phi b_n W \ket{n})
\ee
where the sum is over half of the basis elements.  Here we also assume both $a_n$ and $b_n$ are real but an extra phase  $\phi=i^{N_m}$, that depends on the precise number of $m$-type $\gamma$'s in $W$ , is also included before the $b_n$.

Now lets consider the correlator 
\bea
\bra{\psi} W \ket{\psi} &=& \sum_{nm}  \bra{m} ( a_m + \phi^* b_m W^\dagger ) W ( a_n + \phi b_n W) \ket{n}) \non \\ 
&=&\sum_n a_n b_n (I \pm I)
\eea
 with the $\pm$ depending on whether $ (\phi W)^2 = \pm I$.  
 
In the cases where $ (\phi W)^2 = - I$ the correlator vanishes.  To see when this occurs we need to consider (1) how many permutations or swaps  are needed to bring all operators in two identical words (each of length $N_w$) together and (2)  how many different $m$-type terms occur.  As one needs precisely $ N_w (N_w+1)/2 $ single swaps to bring all corresponding operators in $W^2$ together and $N_m$ is the number of $m$'s in a word (and $N_p$ is the number of $p$'s such that $N_m +N_p= N_w $ ) we see the general condition for the correlators to vanish is that $N_v=N_w (N_w+1)/2 + N_m$ is odd.  

For the case analysed above where we have $W = m_x m_x'$, we see that $N_w=2$ and $N_m=2$ and therefore $N_v = 5$.  Clearly  then $ m_x \ket{\psi}$ and $m_x' \ket{\psi}$ are orthogonal for an arbitrary real state $\ket{\psi}$.  Although it is not generally true that $p_x \ket{\psi}$ and $m_x \ket{\psi}$ are orthogonal, because the relevant $p_x$ and $m_x$ operators tend to be on opposite sides of the system, this does not necessarily present a problem.  However, if we consider for example terms such as  $p_{x_1} m_{x_2} m_{x_3}$ ,  we see that these types of correlations tend to localise to the same side of the system as $m_{x}$. Furthermore we see from the considerations above, that  the overlap between states $p_{x_1} m_{x_2} m_{x_3} \ket{\psi}$ and $m_x \ket{\psi}$ only vanishes only when $x$ is not equal to $x_2$ or $x_3$.  This means that in the interacting system, any measure of the single-particle cross-correlators $\bra{0_e} m_x \ket{0_o}$ is not independent from contributions from the multi-particle part the Majorana operator.  In appendix \ref{sect:subset}, we discuss the possibility of using a sub-sets of operators to represent the mapping between ground states, placing an emphasis on sets of operators that produce states that are almost orthogonal to each other.

\section{Conclusion}

In this paper we have examined the Majorana zero-energy quasi-particles in an interacting regime.  We have used the fact that the commutator of  the free-system can, in the matrix representation generated by Majorana operators, be written in a block-diagonal-form, such that free-particles are the eigensolutions of the sub-block encoding single particle excitations.  We showed that  interactions will disrupt this block-diagonal structure and force eigen-operators  to be superpositions of operators from other sub-blocks encoding multi-particle transitions. In the case of Majorana zero-modes in an interacting regime, because the interactions fail to lift this degeneracy, we can superpose these zero-energy transitions such that the resulting operator is contained mostly in one of these sub-blocks. The many-body Majorana quasi-particle can be thought of as the particular superposition of zero-energy odd-parity transitions that minimise the excitation number.  

We used this idea to calculate the multi-particle content of the Majorana zero modes in the presence of interactions. Our main observation is that, when interactions are present, the multi-particle content of these Majorana excitations can be significant. In retrospect, this is to be expected  because in the definition of the mode we use all energy eigenstates, and interactions should easily generate mixing between eigenstates of the same parity that are not separated by an energy gap.  At first sight, this may conflict with our intuitive understanding of what a topological phase should be.   However,  as was already noted, in the non-interacting regime the {\em structure} of Majorana mode is also very sensitive to small changes of the system parameters and indeed, in it can be argued that this fluidity is the reason the topological degeneracy is so stable.  

We have noted in the main text that the  $U$-dependencies largely follow from block-diagonal structure of the commutator and the notion of the MBM as the mode which has the lowest weight in the higher $A^{(s)}$-blocks.  It is important to stress here that this block structure is not unique to the p-wave wire model and indeed should follow for any quadratic model.  We should therefore expect, provided a many-body zero mode can be shown to exist,  that a similar scaling of the multi-particle content occurs in other models of topological superconductivity that are extended to include quartic interactions.

The results of the paper may also have consequences for what has been called localisation-assisted quantum-order, \cite{Huse2013} where it has been proposed that qubits based on topological phases, which are simultaneously many-body-localised (MBL), may be better protected against decoherence effects. The methodology outlined here, when interpreted through the results of Refs. \onlinecite{Goldstein2012,Yang2014}, suggests that the reduction of the multi-particle rates would be an indication of this effect.  On this point, for weakly interacting systems at least, it is difficult to see how the overall $U$-dependencies of  Eq. \eqref{eq:rates} could be influenced.  However, it might be the case that the $\alpha$-coefficients themselves can be reduced by the addition of disorder, potentially offering a sharp diagnostic.  Another more speculative possibility is that the rates corresponding Eq. \eqref{eq:rates} only describe the very weakly interacting regime and that the transition to the MBL phase drastically re-orders the structure of the Majorana mode.  We leave further discussions of this point for future work.   

A significant portion of the paper has dealt with the issue of the ground-state correlators and how they are related to the Majorana modes in the presence of interactions. We have pointed out that these correlators cannot be used to reliably infer the multi-particle content of the zero-modes.  On this point, it is legitimate to question the importance of the strong quasi-particle picture.  After all, the experimentally relevant physical quantities will be dominated by the properties of the low energy states alone, and we expect that the multi-particle content of the strong-mode will only have an oblique relationship with actual experimental data.  However, this indirect association has important implications for how we should interpret experiments and in particular implies that we should not be too quick to associate experimental data with the single-particle content of the mode itself. 

\section{Acknowledgements}
I thank Niall Moran, Jiri Vala, Joost Slingerland, Denjoe O'Connor, Alessandro Romito, Frank Pollmann, Paul Fendley, David Huse and Dganit Meidan for helpful and informative discussions.  I am also very grateful to the authors of Refs. \onlinecite{Gangadharaiah2011, Stoudenmire2011, Goldstein2012, Yang2014} for critical examinations of the manuscript, useful suggestions regarding the content of the work, and for a number of important clarifications regarding the existing literature.

\appendix

\section{A brief review of the 1d p-wave lattice model}
\label{sect:Kitaevchain}

We are concerned generally with situations where our system can be written as a sum of 
$
H=H_{Q}  + H_{I}
$
where $H_{Q} $ is quadratic free-fermion Hamiltonian 
\be
H_{Q} =  \frac{1}{2} \sum_{i,j=1}^N A^{(1)}_{ij} \gamma_i \gamma_j 
\ee
for imaginary $A_{ij}$  and the interacting terms is a quartic term of the form
\be
H_{I} = \sum_{ijkl} v_{ijkl} \gamma_i \gamma_j \gamma_k \gamma_l
\ee
where $i,j,k$ and $l$ are from the same local neighbourhood. The constituent components here are the position space Majorana terms defined in term of complex Dirac Fermion operators $c^\dagger$ and $c$:
\bea
\label{eq:Majdef1}
p_n&=&\gamma_{2n-1} = \phantom{i} (c^\dagger_n + c_n^\phd ) \\ \non m_n&=&\gamma_{2n\phantom{-1}} = i (c^\dagger_n - c_n^\phd )
\eea
which obey $\{ \gamma_i ,\gamma_j \}= 2 \delta_{ij}$ and thus $\gamma_i = \gamma_i^\dagger$ and $\gamma_i^2 =I$. 

In the main text the particular quadratic model that we have in mind is the 1d spin-less p-wave superconducting model \cite{Kitaev2001}
\bea
H_Q&=&\frac{i }{2} \sum_j^N \mu_l p_{j} m_ {j}  + \frac{i}{2}\sum_{j=1}^{N-1} (|\Delta| + t)  m_{j} p_{j+1} 
\\  &+&  \frac{i}{2} \sum_{j=1}^{N-1} (|\Delta|-t ) p_{j} m_{j+1} \non
\eea
where we have without loss of generality chosen the phase of the p-wave superconducting pairing potential to be real.  The quartic term we use is of the form 
\be
H_{I} = \frac{U}{8} \sum_j p_j m_j  m_{j+1} p_{j+1} 
\ee
When $|\Delta|>0$ and $|\mu_l| < 2t$ the $H_Q$ system is known to be in a topological phase with a Majorana zero modes exponentially localised at each end of the wire\cite{Kitaev2001}.  We will typically work with $\mu = \mu_l +2t$ so that we identify $\mu=0$ as the bottom of the band and the starting point of the topological phase. The transverse Ising model corresponds to the special case of this model where $U=0$ and $t=\Delta$.

 \section{ A spin representation for the $\Gamma$ basis}
 \label{sect:spinGamma}
 
  The matrix $\mathcal{H}$  scales as $4^N \times 4^N$. For $N>8$ this becomes something that is difficult to store on a computer, even if sparse matrix technology is employed.  Ideally we would like to be able to represent this matrix in a more abstract fashion as a sum over operators, in analogy with the way we generate the usual Hamiltonian as a sum over operators that act on basis-states in a well defined way. 
 
We will show how to build our Hilbert space using the $\Ket{\Gamma_a}$ eigenbasis and the $\Ket{\Sigma}$ basis.  One of the nice things about the $\gamma$ operators in general is that they are non-projective , we therefore have on a single site 
 \bea
 m_n  I  &=&  m_n \\
 p_n  I  &=&  p_n \\
p_n p_n =m_n m_n &=& I\\
i Z_n  &=&  m_n p_n  \\
m_n  (i Z_n) &=&  p_n\\
p_n  (iZ_n) &=& -  m_n 
\eea
which means that when acting to the right on a basis defined by  
\bea 
\label{eq:Gamma_basis}
\Ket{I\;\;}&=& \Ket{00}=[1,0]^T \otimes [1,0]^T= [1,0,0,0]^T,\\
\Ket{\;p^\phd}&=&\Ket{01}=[1,0]^T \otimes [0,1]^T=[0,1,0,0]^T,  \non \\ 
\Ket{ m\;}&=&\Ket{10}=[0,1]^T \otimes [1,0]^T=[0,0,1,0]^T, \non \\
 \Ket{iZ}&=&\Ket{11}=[0,1]^T \otimes [0,1]^T=[0,0,0,1]^T \non
\eea
we have
\be
\bar{p\;}^{R}=\left[ \begin{array}{cccc}  0 & \phantom{-}1 & \phantom{-}0 & \phantom{-}0 
\\ 1 &\phantom{-}0 & \phantom{-}0 & \phantom{-}0
\\ 0&  \phantom{-}0&\phantom{-}0 &-1 
\\ 0 &\phantom{-}0 &-1 &\phantom{-}0 \end{array} \right]  = \sigma^z \otimes \sigma^x
\ee
\be
\bar{m\;}^{R}=\left[ \begin{array}{cccc} 0 &  \phantom{-}0 &  \phantom{-}1 &  \phantom{-}0 
\\ 0 &  \phantom{-}0 &  \phantom{-}0 &  \phantom{-}1
\\ 1&   \phantom{-}0&  \phantom{-}0 & \phantom{-}0 
\\ 0  &  \phantom{-}1 & \phantom{-}0 & \phantom{-}0 \end{array} \right] = \sigma^x \otimes I_2
\ee

To enforce anti-commutation relations between different sites we need to attach  Jordan-Wigner (JW) like strings which should anti-commute with $m^{R}$ and $p^{R}$ but take all  basis elements onto themselves.  In the above basis, one such  operator is 
\be
S = \left[ \begin{array}{cccc} 1 & 0 & 0 & 0 \\ 0 & -1 & 0 & 0\\  0&  0 & -1 &0 \\ 0 & 0 & 0 & 1 \end{array} \right] = \sigma^z \otimes \sigma^z.
\ee
and we then have
\bea
m^{R}_x =[ \prod_{j=1}^{x-1} S_j ] \times \bar{m}^{R}_x, \quad p^{R}_x = [\prod_{j=1}^{x-1} S_j ] \times \bar{p \;}^{R}_x \\
\eea

Similarly for the action to the left can write have
\be
\bar{p\;}^{L}=\left[ \begin{array}{cccc}  0 & \phantom{-}1 & \phantom{-}0 & \phantom{-}0 
\\ 1 &\phantom{-}0 & \phantom{-}0 & \phantom{-}0
\\ 0&  \phantom{-}0&\phantom{-}0 &\phantom{-}1 
\\ 0 &\phantom{-}0 &\phantom{-}1 &\phantom{-}0 \end{array} \right]=I_2 \otimes \sigma^x
\ee
\be
\bar{m}^{L}=\left[ \begin{array}{cccc} 0 &  \phantom{-}0 &  \phantom{-}1 &  \phantom{-}0 
\\ 0 &  \phantom{-}0 &  \phantom{-}0 &  -1
\\ 1&   \phantom{-}0&  \phantom{-}0 & \phantom{-}0 
\\ 0  &  -1 & \phantom{-}0 & \phantom{-}0 \end{array} \right]=  \sigma^x \otimes \sigma^z
\ee
 
In order to enforce anti-symmetry of the left-acting operators alone we can choose our JW-strings so that they come from the opposite direction
\bea
m^{L}_x =[ \prod_{j=x+1}^{N_x} S_j ] \times \bar{m}^{L}_x, \quad p^{L}_x = [\prod_{j=x+1}^{N_x} S_j ] \times \bar{p \;}^{L}_x \non
\eea
However, this choice means that operators from the left and right commute.  This does not have to be the case and we can also choose our J-W strings so that all operators anti-commute. One such choice would be to to set
\bea
\label{eq:gamma_mpR}
m^{R}_x =[ \prod_{j=1}^{x-1} S_j ] \times \bar{m}^{R}_x, \quad p^{R}_x = [\prod_{j=1}^{x-1} S_j ] \times \bar{p \;}^{R}_x  \non
\eea
and
\bea
\label{eq:gamma_mpL}
m^{L}_x =i[ \prod_{j=1}^{x} S_j ] \times \bar{m}^{L}_x, \quad p^{L}_x = -i[\prod_{j=1}^{x} S_j ] \times \bar{p \;}^{L}_x \non
\eea
where the additional $i$ phases are chosen so that $m^2 = p^2 =I$. Note that there is some freedom in the choice of overall sign here which can be useful for switching the overall sign of $H_L$ for example. 
 
\section{A spin basis for Dirac fermions }
\label{sect:spinSigma}
\begin{figure}
\includegraphics[width=0.4\textwidth,height=0.3\textwidth]{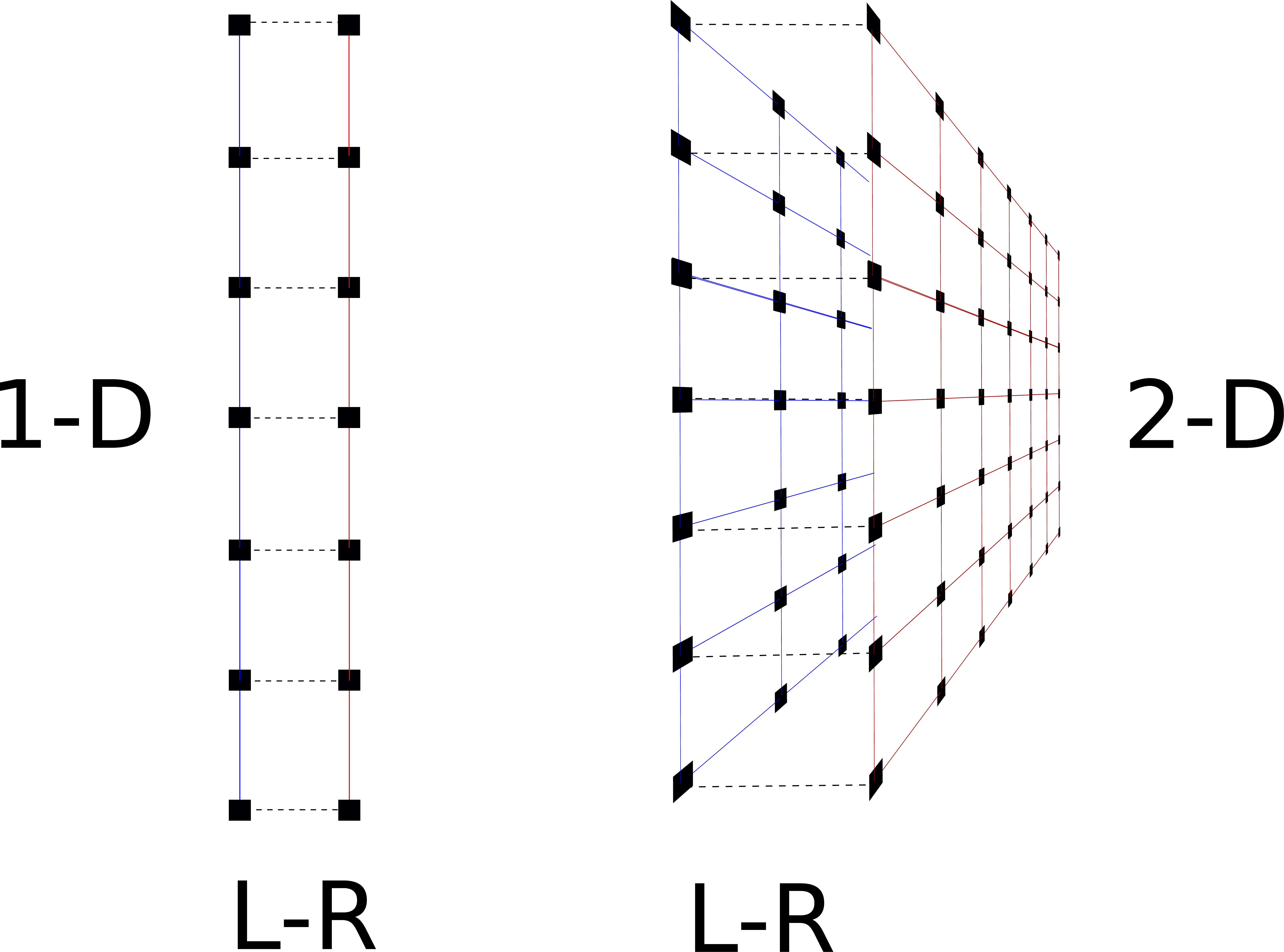}
\caption{Fermionic doubling: in the $\Sigma$-basis, the commutator $[H,\bullet]$ can be understood as two disconnected copies of original system. }
 \label{fig:DoublingFigure}
 \end{figure}
In the previous section we showed how to build up a representation for the $\Ket{\Gamma}$ basis which is based on the properties of the Clifford algebra. However there are other ways to do this.  Consider the Fock space representation for a single two level mode we have
\bea
\label{eq:sigmarules}
\sigma^+ &=& +\ket{1}\bra{0}  \\
\sigma^-&=& +\ket{0}\bra{1} \non \\
\sigma^- \sigma^+ &=& +\ket{0}\bra{0} \non \\
\sigma^+ \sigma^- &=& +\ket{1}\bra{1} \non \\
\sigma^- \sigma^+ +\sigma^+ \sigma^-= I_2 &=&  +\ket{0}\bra{0} +\ket{1}\bra{1} \non \\
\sigma^z &=&-\ket{0}\bra{0} + \ket{1}\bra{1} \non
\eea
 
Now for a single site we choose
\bea
\Bra{\Sigma^1} &= & \Bra{\sigma^- \sigma^+ } = [1,0,0,0] \\
\Bra{\Sigma^2} &=& \Bra{\;\;\; \sigma^+ \phd} = [0,1,0,0]\\
\Bra{\Sigma^3} &=& \Bra{\;\;\; \sigma^- \phd}=[0,0,1,0] \\
\Bra{\Sigma^4} &=& \Bra{\sigma^+ \sigma^-  }=[0,0,0,1] 
\eea
To represent each operator in this basis we need to see how the operators act to both the right (on $\Ket{\Sigma}$ states) and the left (on $\Bra{\Sigma}$ states). As will be seen, it is enough to examine  $\sigma^-$ in each scenario. Acting to the right the $\sigma^-$ operator should send $\Ket{\sigma^+ \sigma^- } \rightarrow \Ket{\sigma^-} $ and $\Ket{\sigma^+} \rightarrow \Ket{\sigma^- \sigma^+ }  $ . Therefore we have

\be
\sigma^-_R=\left[ \begin{array}{cccc} 0 & 1 & 0 & 0 \\ 0 & 0 & 0 & 0\\  0&  0&0 &1 \\0 &0 &0 &0 \end{array} \right]
\ee

On the other hand, when acting to the left , $\sigma^-$ should send $\Bra{\sigma^+ \sigma^-} \rightarrow \Bra{\sigma^+} $ and $\Bra{\sigma^- } \rightarrow \Bra{\sigma^- \sigma^+} $ .  Recall that when operating to the left, the conjugate of the operating term appears inside the left-hand-side basis state , but to the right of the existing operator label. Therefore we write
\be
\sigma^-_L=\left[ \begin{array}{cccc} 0 & 0 & 0 & 0 \\ 0 & 0 & 0 & 0\\  1&  0&0 &0 \\0 &1 &0 &0 \end{array} \right]
\ee

The importance of this particular basis becomes clear when we examine the $\pm$ superposition and we have
\bea
X^R &=&  \phantom{i(}\sigma^+_R +\sigma^-_R\phantom{)}=
\left[ \begin{array}{cccc} 0 & \phantom{-}1 & \phantom{-}0 & \phantom{-}0 \\ 
1 & \phantom{-}0 & \phantom{-}0 & \phantom{-}0\\ 
0&  \phantom{-}0& \phantom{-}0 &\phantom{-}1 \\ 
0 &\phantom{-}0 &\phantom{-}1 &\phantom{-}0 \end{array} \right] = I \otimes \sigma^x \non
\eea
\bea
X^L&=& \phantom{i(} \sigma^+_L +\sigma^-_L\phantom{)}=
\left[ \begin{array}{cccc} 0 & \phantom{-}0 & \phantom{-}1 & \phantom{-}0 \\ 
0 & \phantom{-}0 & \phantom{-}0 & \phantom{-}1\\ 
1&  \phantom{-}0& \phantom{-}0 &\phantom{-}0 \\ 
0 &\phantom{-}1 &\phantom{-}0 &\phantom{-}0 \end{array} \right] = \sigma_x \otimes I \non
\eea
\bea
Y^R&=& i (\sigma^+_R -\sigma^-_R)=
\left[ \begin{array}{cccc} 0 & -i & \phantom{-}0 & \phantom{-}0\\ 
i & \phantom{-}0 & \phantom{-}0 & \phantom{-}0\\
0 &  \phantom{-}0& \phantom{-}0 & -i \\ 
0 &\phantom{-}0 &\phantom{-}i &\phantom{-}0\end{array} \right]=I \otimes \sigma^y \non
\eea
\bea
Y^L&=&i (\sigma^+_L -\sigma^-_L)=
\left[ \begin{array}{cccc} \phantom{-}0 & \phantom{-}0 & \phantom{-}i & \phantom{-}0 \\ 
\phantom{-}0 & \phantom{-}0 & \phantom{-}0 & \phantom{-}i\\ 
-i &  \phantom{-}0& \phantom{-}0 &\phantom{-}0 \\ 
\phantom{-}0 &-i &\phantom{-}0 &\phantom{-}0 \end{array} \right] = -\sigma_y \otimes I \non
\eea
 To create a fermionic basis we can attach Jordan-Wigner strings and we could write for example the Majorana fermion operators as
\be
p^R_n =  I \otimes  (c^\dagger_n+c^\phd_n )  = [ \prod_{i=1}^{n-1} I \otimes \sigma_i^z ] X_n^R
\ee 
or
\be
m^L_n =   i(c^\dagger_n-c^\phd_n )\otimes I  = [ \prod_{i=1}^{n-1} \sigma_i^z \otimes I ] Y_n^L
\ee 
The $\Ket{\Sigma}$ representation above reveals that left operating operators act on an entirely different sub-space to the right. This means that we can represent any transition Hamiltonian matrix as
 \be
 \mathcal{H} = [H,\bullet] = I \otimes H - H \otimes I = H_R-H_L
 \ee
 and therefore the transition matrix is simply a trivial doubling of the original Hamiltonian but where all constants on one Hamiltonian have been negated.  Similar observations with respect to integrability have been made in the context of parafermions. \cite{Fendley2012}  For a 1-d and 2-d  systems this lends itself to the easy visualisation shown in Figure \ref{fig:DoublingFigure}. 
 
We are free to interpret the transition Hamiltonian as (i) two separate fermionic systems where the fermions of left and right do not anti-commute with each other or (ii) as a single fermionic system which as no terms than connect sites with index $L$ to  sites with index $R$.  This can easily be achieved in the construction above by choosing a Jordan-Wigner string convention that runs through both indices $L$ and $R$. In this latter picture the opposite overall sign on the $H_L$ terms can make interpretation slightly more cumbersome, in particular for lattice system where we would like to take the continuum limit. However,  we note that the trivial transformation $c_L^\dagger \leftrightarrow c_L$ sends  $m_L \rightarrow -m_L$ and therefore any left-hand terms with an odd number of $m$-type operators will also change sign under this change of basis. 
 
This $\Ket{\Sigma}$ basis is related to the $\Ket{\Gamma}$ basis by Hadamard rotations  from the $p$ and $m$  to the $c^\dagger$ and $c^\phd$  together with additional Hadamard rotations from  $Z$ and $I$  to the $c^\dagger c^\phd$ and $c^\phd c^\dagger$.  The first transformation is what we understand on the single particle level as a change of basis from a Bogoliubov de-Gennes representation to a Kastelyn-like Majorana adjacency representation. These rotations take place entirely within each of the sub-blocks  $A^{(n)}$. By contrast, the second transformation mixes between different $A^{(n)}$ sub-blocks. In section \ref{sect:Quadratic} we show that this transformation is essential to understanding relationship between solutions of each sub-block and actual eigensolutions of the full commutator $\ket{n}\bra{m}$.

\section{Review of full-diagonalisation method (FD) for computing zero-energy Majorana modes in the presence of local interactions}
\label{sect:Algorithm_fulldiag}

In Ref. \onlinecite{Kells2014} it was demonstrated that , in the $L \gg \xi$ limit, there was a well-defined notion of the Majorana quasi-particle even in the presence of strong interactions.  The stability of the Majorana to strong interactions follows from the fact that the degeneracy between all even-odd pairs remains to an order of perturbation theory that scales with the length of the system. This degeneracy then allows one to calculate the precise structure of the Majorana modes by:

\noindent(1) Calculating all eigenfunctions of even and odd sectors. \\ \\
(2) Confirm even-odd counterparts by checking for example that $\bra{n_o} \gamma_{L}(U=0) \ket{n_e}$ is large. \\ \\
(3) Fixing the relative phases of all even-odd pairs using the bare-non interacting Majorana modes. For the situation with real coefficients only we calculate $
s_n^{(R)}=\text{sign} (\bra{n_o} \beta_1^\dagger + \beta_1 \ket{n_e})
$ and  set $\ket{n_o} \rightarrow s^{(R)}_n \ket{n_o}$ . 
\\ \\
\noindent (4) Finally  with $s_n^{(L)}=\text{sign} (\bra{n_o} \beta_1^\dagger - \beta_1 \ket{n_e})$, we can then write 
\bea
\label{eq:gamma_bar}
\gamma_R &=& \phantom i \sum \;\;I \;\; \ket{n_o}  \bra{n_e}  + \;\; I \;\; \ket{n_e} \bra{n_o}  \\
\gamma_L &=& i \sum  s^{(L)}_n \ket{n_o}  \bra{n_e}  - s^{(L)}_n \ket{n_e}  \bra{n_o} .
\non 
\eea

 \begin{figure}
\includegraphics[width=0.45\textwidth,height=0.3\textwidth]{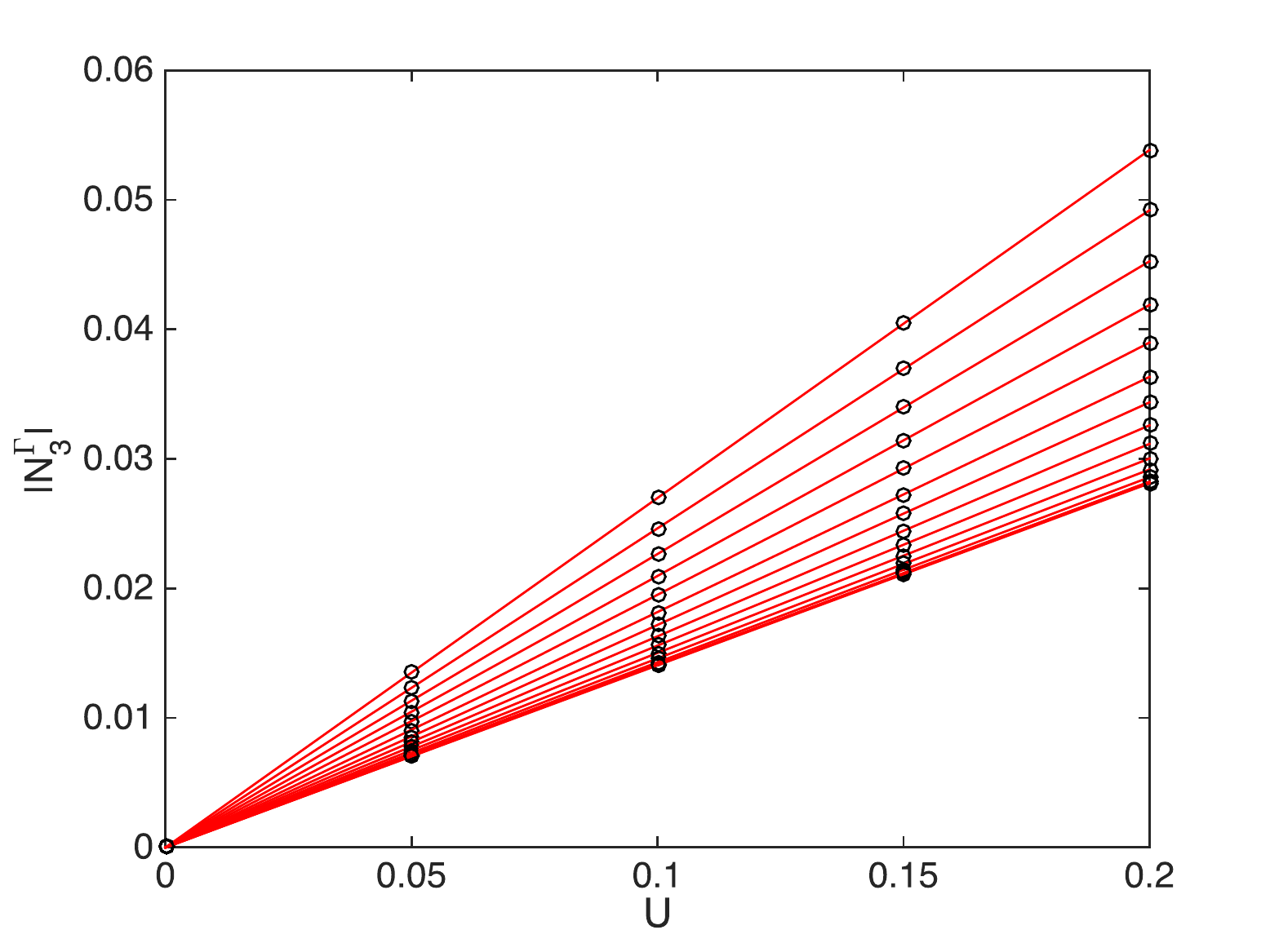}
\caption{In the figure we show how $|N_3^\Gamma|$ grows for a fixed value of $\Delta =0.8$ and different values of $\mu \in [.7 , 2.1]$. The values of $\alpha_3$ at different parameters represent the slopes  of these straight lines.  A system size of $N_x=50$, with cut-offs $(N_s,N_d)=(9,9)$ was used for this plot.}
 \label{fig:linear_N3}
 \end{figure}

 \begin{figure}
\includegraphics[width=0.45\textwidth,height=0.3\textwidth]{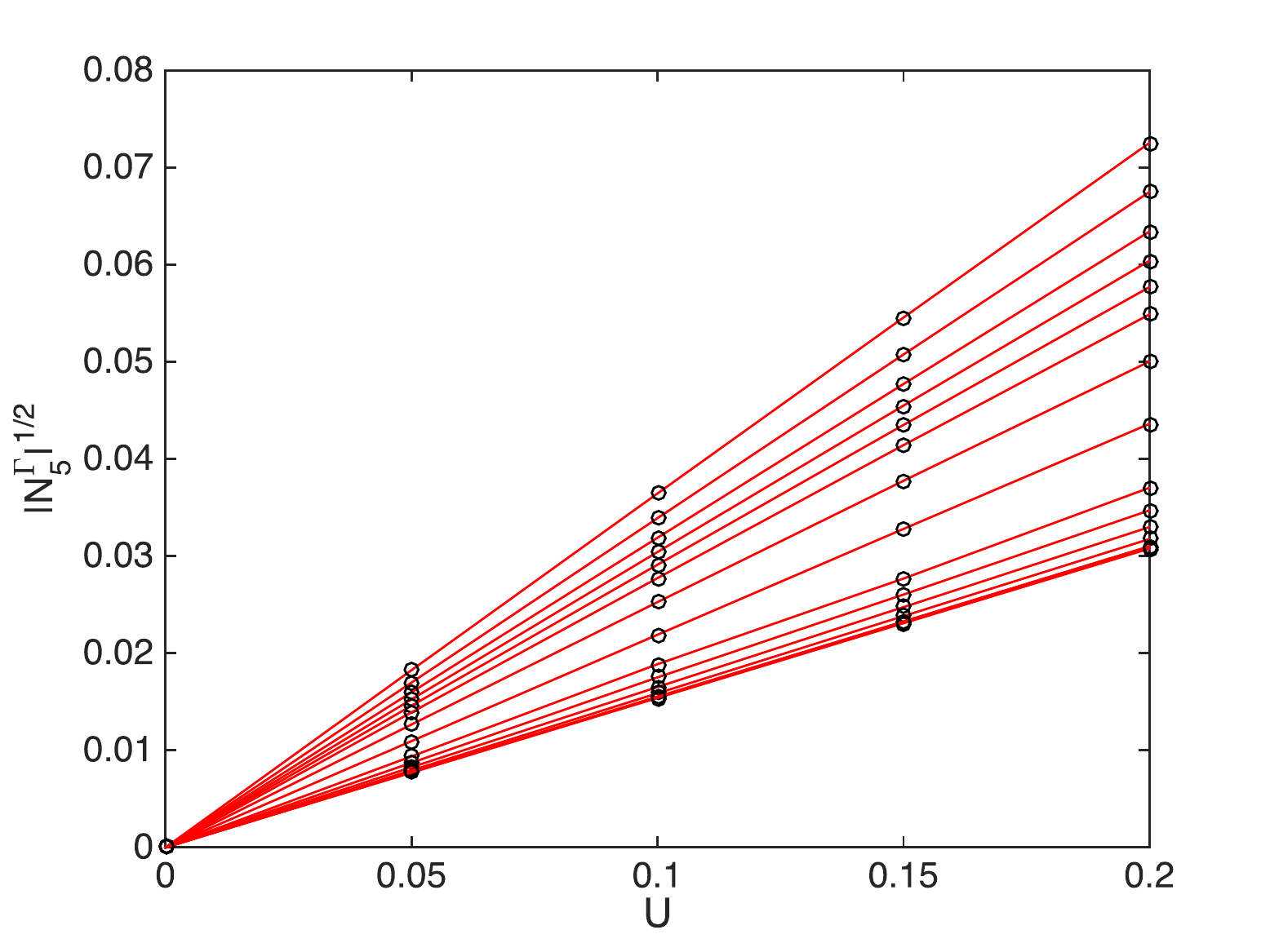}
\caption{In the figure we show how $|N_5^\Gamma|^{1/2}$ grows for a fixed value of $\Delta =0.7$ and different values of $\mu \in [.7 , 2.1]$. The values of $\alpha_5^{1/2}$ at different parameters represent the slopes of these straight lines.  A system size of $N_x=50$, with cut-offs $(N_s,N_d)=(9,9)$ was used for this plot.}
 \label{fig:linear_N5}
 \end{figure}

 \begin{figure}
\includegraphics[width=0.45\textwidth,height=0.3\textwidth]{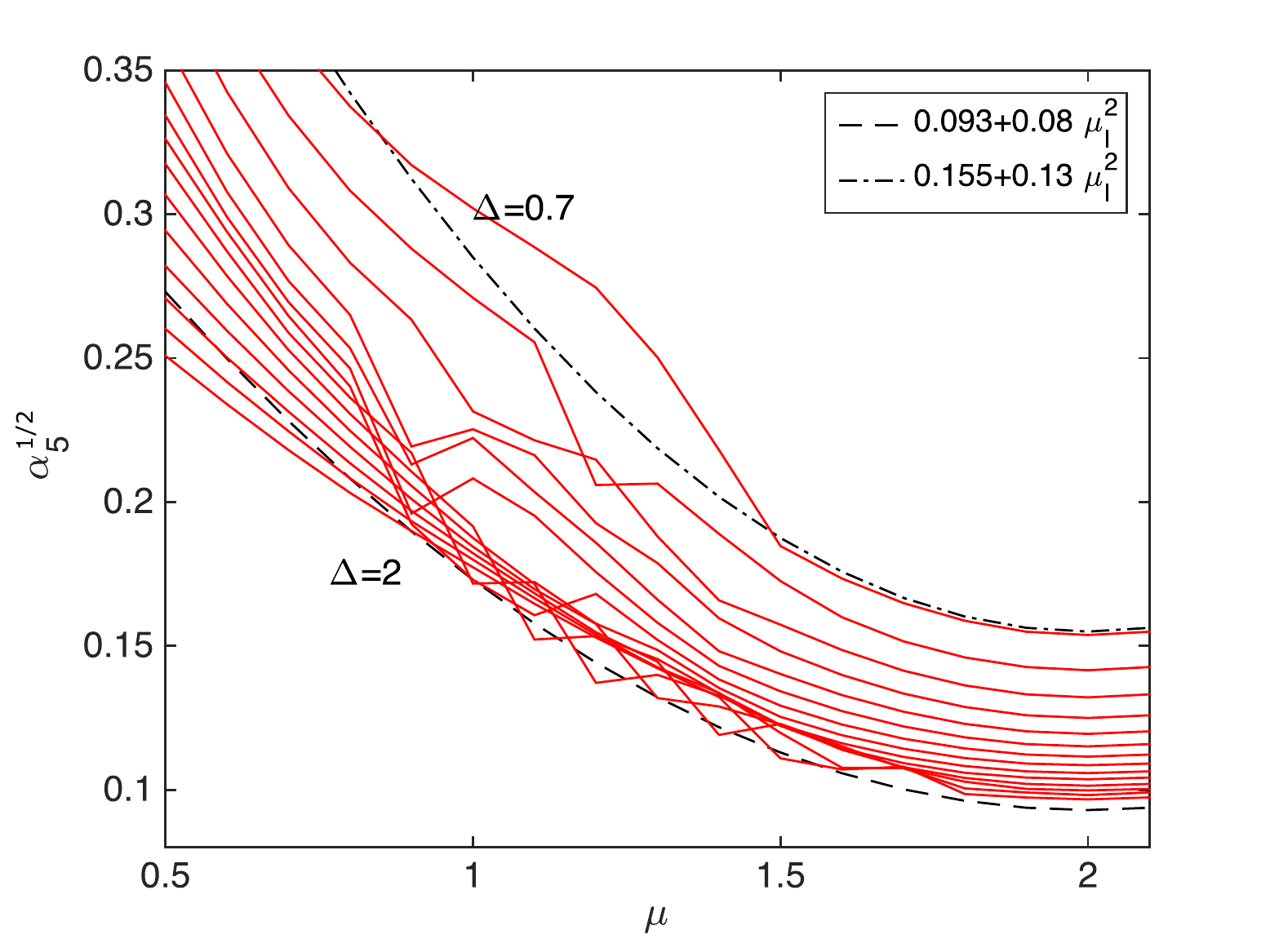}
\caption{In the figure we plot the rate $\alpha_5^{1/2}$ as a function of $\mu$. We again see a quadratic dependence about the linearised dispersion point at $\mu=2$.   The plot shows some numerical instability. Note that to fit these curves we take the $|N_5|^{1/4}|$ and very small numerical errors get magnified to some degree.   A system size of $N_x=50$, with cut-offs $(N_s,N_d)=(9,9)$ was used for this plot. }
 \label{fig:alpha5_muplot}
 \end{figure}

\begin{figure}
\includegraphics[width=0.45\textwidth,height=0.3\textwidth]{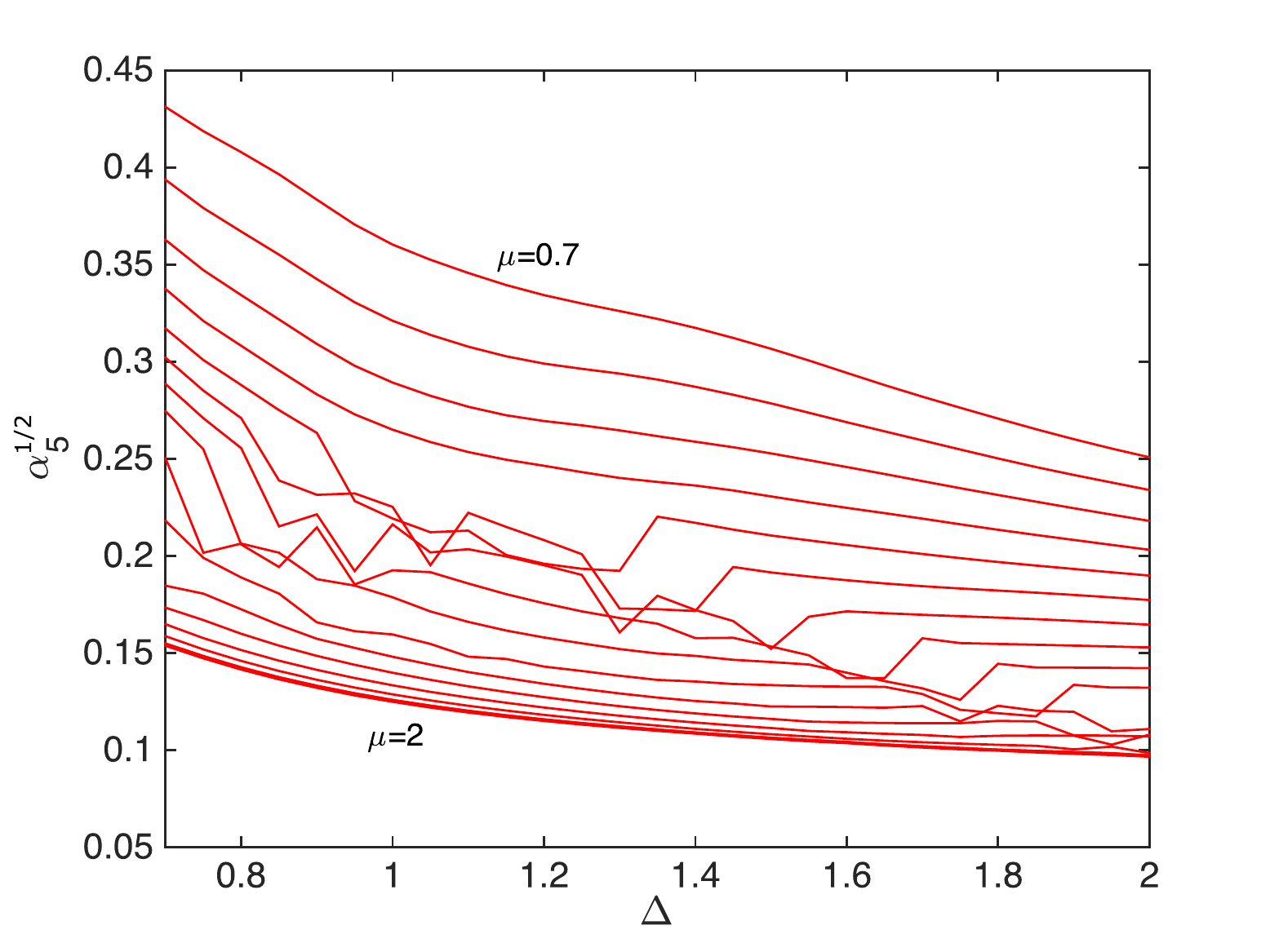}
\caption{In the figure we plot the rate $\alpha_5^{1/2}$ as a function of $\Delta$.  A system size of $N_x=50$, with cut-offs $(N_s,N_d)=(9,9)$ was used for this plot.  }
 \label{fig:alpha5_deltaplot}
 \end{figure}

\section{Additional Numerical results }
\label{sect:AddNum}

In this section we provide additional numerics which give further support for the central claims of the main text regarding the N-particle content of the Majorana zero modes.  In Figure \ref{fig:linear_N3} we clearly see how $|N_3^\Gamma|$ depends linearly on $U$ for a variety of system parameters.  A similar story is also evident in Figure \ref{fig:linear_N5} where we plot $|N_5^\Gamma|^{1/2}$ for a variety of system parameters. Although the linear dependence on $U$ is clear we see some apparent fluctuations in how the slope changes for different values of $\mu$.  

We plot the values of $\alpha^{1/2}_5$ directly in Figures \ref{fig:alpha5_muplot} and \ref{fig:alpha5_deltaplot}.  We note that $\alpha_5^{1/2}$ has the same general dependence on $\mu$ and $\Delta$ as $\alpha_3$ .  However we also see that the fluctuations in $\alpha_5^{1/2}$ are not specific to the parameters used for Figure \ref{fig:linear_N5} above.  We suspect that the fluctuations are probably a numerical artefact resulting from the finite cut-offs mentioned in section \ref{sect:Algorithms}. Nonetheless, since we have not been able to remove this effect by going to larger system sizes (and larger cut-offs) we cannot discount the possibility that it is due to some unknown physical effect.

\section{Complete operator sets for mapping between ground states}
\label{sect:subset}

In section \ref{sect:correlators} we argued that ground state correlation data cannot be used to infer the structure of the Majorana mode in the interacting regime. This is because the problem of finding an operator such that $ | \bra{0_e} O \ket{0_o}| =1$ is under-defined, and there are many different ways of satisfying this criteria.  Nonetheless, it is still meaningful to try to find a sensible expansion (in position space operators $\Gamma_a$) of an operator $O$ that fits the aforementioned criteria. 

Ideally we would like to find a sets of $2^{N-1}$ operators  that send some arbitrary state $\ket{\psi}$ (i.e one of the ground states ) to an orthonormal basis of states in the other sector. This would allow us to uniquely capture the structure of the two Majorana zero-mode operators which connect ground states in even(odd) sectors.   We find that it is relatively easy to find a set of operators that gives an independent basis.  Orthogonality on the other hand, while achievable, requires one to employ a procedure such as Gram-Schmidt and the resulting operators end up being complicated superpositions which do not allow us to extract any physical intuition.

Although orthogonality is not practical, it is possible to come up with a set of operators that generates a partially orthonormal basis. This then does allow us some intuitive understanding  of the Majorana zero-mode structure in the sense implied by Ref. \onlinecite{Stoudenmire2011}. This operator choice  for the p-wave system  is outlined in Table \ref{table:LRcorrelators}. An appealing property of this choice is that the $ppp$-type operators are always `orthogonal' to $m$-type in the sense that for any state $\ket{\psi}$ we have $\bra{\psi} m_{x_1} p_{x_2} p_{x_3} p_{x_4}  \ket{\psi}=0$. Indeed, each sub-set of operators picks out states that are always orthogonal to the states picked out by the set immediately above or below them in the table.  The problem arises however in that, apart from the single particle rows $\Gamma^{(1)}$, each set of operators is not orthogonal with elements of the same type when these operators overlap on one common site, see section \ref{sect:correlators}.  By the same reasoning,  members of the set $mmmmm$  overlap with members of $m$, and states obtained by operating with $ppp$ will have to overlap with some members of $ppppppp$ etc. Therefore, even with this carefully chosen set of operators, the condition $ | \bra{0_e} O \ket{0_o}| =1$ does not appear to be restrictive enough to define a unique operator.  

\begin{table}[ht]
\centering
\begin{tabular}{ c || c | c  }
\hline
    & Left-localised & Right-localised  \\ \hline 
    $\Gamma^{(1)}$  &$m$ & $p$   \\ 
    $\Gamma^{(3)}$  & $p_\phd p_\phd p_\phd $ & $mmm$   \\
    $\Gamma^{(5)}$  & $mmmmm$ & $p_\phd p_\phd p_\phd p_\phd p_\phd$ \\
    $\Gamma^{(7)}$  & $p_\phd p_\phd p_\phd p_\phd p_\phd p_\phd p_\phd $ & $mmmmmmm$ \\
     & \vdots & \vdots
\end{tabular}
\caption{A set of $2n-1$ point correlators that can be used to describe the many-body Majorana operators. In the table the $m$ stands for $m_x$, $ppp$ stands for $p_{x_1} p_{x_2} p_{x_3}$ etc.  where $m_x = i (c^\dagger_x-c_x) $ and $p_x = c_x^\dagger+c_x$ . }
\label{table:LRcorrelators}
\end{table}

\end{document}